\documentclass[10pt,journal,onecolumn]{IEEEtran}
\usepackage{epsfig}
\usepackage{amsmath}
\usepackage{amstext}
\usepackage{amsfonts}
\usepackage{amssymb}
\usepackage{eucal}
\usepackage{graphicx}
\usepackage{verbatim}
\usepackage{stmaryrd}
\usepackage{pstricks}
\usepackage{pst-node}
\usepackage{psfrag}

\usepackage{algorithmic}
\usepackage{booktabs}

\newcommand{\RR}{{\mathbb{R}}}

\newcommand{\CC}{{\mathbb{C}}}

\newcommand{\EE}{{\rm E}}
\newcommand{\B}{{\bf B}}
\newcommand{\D}{{\bf D}}
\newcommand{\Z}{{\bf Z}}
\newcommand{\E}{{\bf E}}
\newcommand{\A}{{\bf A}}

\newcommand{\X}{{\bf X}}
\newcommand{\Y}{{\bf Y}}
\renewcommand{\S}{{\bf S}}
\renewcommand{\P}{{\bf P}}
\newcommand{\T}{{\bf T}}

\newcommand{\U}{{\bf U}}
\renewcommand{\O}{{\bf O}}
\newcommand{\R}{{\bf R}}
\newcommand{\Q}{{\bf Q}}
\renewcommand{\H}{{\bf H}}
\newcommand{\I}{{\bf I}}
\newcommand{\x}{{\bf x}}

\newcommand{\s}{{\bf s}}

\newcommand{\y}{{\bf y}}

\newcommand{\n}{{\bf n}}
\newcommand{\w}{{\bf w}}
\newcommand{\e}{{\bf e}}

\renewcommand{\b}{{\bf b}}
\newcommand{\oh}{{\frac{1}{2}}}
\newcommand{\sgn}{{\rm sgn}}
\newcommand{\asto}{\overset{\rm a.s.}{\longrightarrow}}

\newcommand{\rv}{{\bf r}}

\newcommand{\ui}{\underline i}

\newcommand{\herm}{{\sf H}}
\newcommand{\rank}{{\rm rank}}
\newcommand{\trans}{{\sf T}}

\newcommand{\C}{{\mathcal C}}
\newcommand{\CCC}{{\bf C}}

\newcommand{\MAC}{{\rm MAC}}

\DeclareMathOperator{\tr}{tr}
\DeclareMathOperator{\diag}{diag}

\newtheorem{theorem}{Theorem}
\newcounter{cproposition}
\newtheorem{proposition}[cproposition]{Proposition}
\newcounter{ccorollary}
\newtheorem{corollary}[ccorollary]{Corollary}
\newcounter{clemma}
\newtheorem{lemma}[clemma]{Lemma}

\newcounter{cremark}
\newtheorem{remark}[cremark]{Remark}

\begin{document}
\bibliographystyle{IEEEtran}

\title{A Deterministic Equivalent for the Analysis of Correlated MIMO Multiple Access Channels}

\author{Romain~Couillet,~\IEEEmembership{Student Member,~IEEE,}~M{\'e}rouane~Debbah,~\IEEEmembership{Senior Member,~IEEE}~and~Jack~W.~Silverstein
  \thanks{R. Couillet and M. Debbah are with the Alcatel-Lucent Chair on Flexible Radio,
    SUP{\'E}LEC, Gif sur Yvette, 91192, Plateau de Moulon, 3, Rue
    Joliot-Curie, France e-mail: \{romain.couillet,~merouane.debbah\}@supelec.fr. M. Debbah's work is supported by the European Commission, FP7 Network of Excellence in Wireless Communications NEWCOM++ and the French ANR Project SESAME.}
  \thanks{J. W. Silverstein is with the Department of Mathematics, North Carolina State University, Raleigh, North Carolina 27695-8205, jack@math.ncsu.edu. J. W. Silverstein's work is supported by the U.S. Army Research Office, Grant W911NF-09-1-0266.}
}

\maketitle

\begin{abstract}
In this article, novel deterministic equivalents for the Stieltjes transform and the Shannon transform of a class of large dimensional random matrices are provided. These results are used to characterise the ergodic rate region of multiple antenna multiple access channels, when each point-to-point propagation channel is modelled according to the Kronecker model. Specifically, an approximation of all rates achieved within the ergodic rate region is derived and an approximation of the linear precoders that achieve the boundary of the rate region as well as an iterative water-filling algorithm to obtain these precoders are provided. An original feature of this work is that the proposed deterministic equivalents are proved valid even for strong correlation patterns at both communication sides. The above results are validated by Monte Carlo simulations.
\end{abstract}

\begin{IEEEkeywords}
Deterministic equivalent, random matrix theory, ergodic capacity, MIMO, MAC, optimal precoder.
\end{IEEEkeywords}

\section{Introduction}
\label{sec:intro}
When mobile networks were expected to run out of power and frequency resources while being simultaneously subject to a demand for higher transmission rates, Foschini \cite{FOS98} introduced the idea of multiple input multiple output (MIMO) communication schemes. Telatar \cite{TEL99} then predicted a growth of the channel capacity by a factor $\min(N,n)$ for an $N\times n$ MIMO system compared to the single-antenna case when the matrix-valued channel is modelled with independent and identically distributed (i.i.d.) standard Gaussian entries. In practical systems though, this linear gain can only be achieved for high signal-to-noise ratios (SNR) and for uncorrelated transmit and receive antenna arrays at both communication sides. Nevertheless, the current scarcity of available frequency resources has led to a widespread incentive for MIMO communications. Mobile terminal engineers now embed numerous antennas in small devices. Due to space limitations, this inevitably induces channel correlation and thus reduced transmission rates. An implication of the results introduced in this paper is the ability to study the performance of MIMO systems subject to strong correlation effects in multi-user and multi-cellular contexts, a question which is paramount to cellular service providers.

Although alternative communication models could be treated using similar mathematical expressions, such as cooperative and non-cooperative multi-cell communications with users equipped with multiple antennas, the present article investigates the MIMO multiple access channel (MIMO-MAC), where $K$ multi-antenna mobile terminals transmit information to a single receiver. Under perfect channel state information at the transmitters (CSIT), the boundaries of the achievable rate region of the MIMO-MAC have been characterised by Yu {\it et al.} \cite{CIO04} who provide an iterative water-filling algorithm to obtain the sum rate maximising precoders. However, to achieve perfect CSIT, the channel must be quasi-static during a sufficiently long period to allow feedback or pilot signalling from the receiver to the transmitters. For high mobility wireless services, this is often unacceptable. In this situation, the transmitters are often assumed to have statistical information about the random fast varying channels, such as first order moments of their distribution. The achievable rates are in this case the points lying in the {\it ergodic} rate region. It is however difficult to characterise the boundary of the ergodic rate region because it is difficult to compute the optimal precoders that reach the boundaries. In the single-user context, an algorithm was provided by Vu {\it et al.} in \cite{VU05} to solve this problem. However, the technique of \cite{VU05} is rather involved as it requires nested Monte Carlo simulations and does not provide any insight on the nature of the optimal precoders. 

In the present article, we provide a parallel approach that consists in approximating the ergodic sum rate by {\it deterministic equivalents}. That is, for all finite system dimensions, we provide an approximation of the ergodic rates, which is accurate as the system dimensions grow asymptotically large. Furthermore, we provide an efficient way to derive an asymptotically accurate approximation of the optimal precoders. The mathematical field of large dimensional random matrices is particularly suited for this purpose, as it can provide deterministic equivalents of the achievable rates depending only on the relevant channel parameters, e.g., the long-term transmit and receive channel covariance matrices in the present situation, the deterministic line of sight components in Rician models as in \cite{HAC07} etc. The earliest notable result in line with the present study is due to Tulino {\it et al.} \cite{TUL05}, who provide an expression of the asymptotic ergodic capacity of point-to-point MIMO communications when the random channel matrix is composed of i.i.d. Gaussian entries. 
In \cite{PEA08}, Peacock {\it et al.} extend the asymptotic result of \cite{TUL05} in the context of multi-user communications by considering a $K$-user MAC with channels $\H_1,\ldots,\H_K$ modelled as Gaussian with a separable variance profile. This is, the entries of $\H_k$ are Gaussian independent with $(i,j)$-th entry of zero mean and variance $\sigma^2_{k,ij}$ that can be written as a product $\sigma^2_{k,ij}=r_{k,i}t_{k,j}$ of a term depending on $i$ and a term depending on $j$. The asymptotic eigenvalue distribution of this matrix model is derived, but neither any explicit expression of the sum rate is provided as in \cite{TUL05}, nor is any ergodic capacity maximising policy derived. In \cite{SOY09}, Soysal {\it et al.} derive the sum rate maximising precoder policy in the case of a MAC channel with $K$ users whose channels $\H_1,\ldots,\H_k$ are one-side correlated zero mean Gaussian, in the sense that all rows of $\H_k$ have a common covariance matrix, different for each $k$. 

In this article, we concentrate on the more general {\it Kronecker} channel model. This is, we assume a $K$-user MIMO-MAC, with channels $\H_1,\ldots,\H_K$, where each $\H_k$ can be written in the form of a product $\R_k^\oh\X_k\T_k^\oh$ where $\X_k$ has i.i.d. zero mean Gaussian entries and the left and right correlation matrices $\R_k$ and $\T_k$ are deterministic nonnegative definite Hermitian matrices. This model clearly covers the aforementioned channel models of \cite{TUL05}, \cite{PEA08} and \cite{SOY09} as special cases. The Kronecker model is particularly suited to model communication channels that show transmit and receive correlations, different from one user to the next, in a rich scattering environment. 
Nonetheless, the Kronecker model is only valid in the absence of a line-of-sight component in the channel, when a sufficiently large number of scatterers is present in the communication medium to justify the i.i.d. aspect of the inner Gaussian matrix and when the channel is frequency flat over the transmission bandwidth. Using similar tools as those used in this article, many works have studied these channel models, mostly in a single-user context. We remind the main contributions, from which the present work borrows several ideas. In \cite{HAC07,DUM10,HAC08}, Hachem {\it et al.} study the point-to-point multi-antenna Rician channel model, i.e., non-central Gaussian matrices with a variance profile, for which they provide a deterministic equivalent of the ergodic capacity \cite{HAC07}, the corresponding ergodic capacity-achieving input covariance matrix \cite{DUM10} and a central limit theorem for the ergodic capacity \cite{HAC08}. In \cite{MOU07}, Moustakas {\it et al.} provide an expression of the mutual information in time varying frequency selective Kronecker channels, using the replica method \cite{MEZ88}. This result has been recently proved by Dupuy {\it et al.} in a yet unpublished work. Dupuy {\it et al.} then derived the expression of the capacity maximising precoding matrix for the frequency selective channel \cite{DUP10}. A more general frequency selective channel model with non-separable variance profile is studied in \cite{RAS08} by Rashibi {\it et al.} using alternative tools from free probability theory. Of practical interest is also the theoretical work of Tse \cite{CHU02} on MIMO point-to-point capacity in both uncorrelated and correlated channels, which are validated by ray-tracing simulations.

The main contribution of this paper is summarized in two theorems contributing to the field of random matrix theory and enabling the evaluation of the ergodic rate region of the MIMO-MAC with Kronecker channels. We subsequently derive an iterative water-filling algorithm enabling the description of the boundaries of the rate region by providing an expression of the asymptotically optimal precoders.
The remainder of this paper is structured as follows: in Section \ref{sec:summary}, we provide a short summary of the main results and how they apply to multi-user wireless communications. In Section \ref{sec:maths}, the two theorems are introduced, the complete proofs being left to the appendices. In Section \ref{sec:capa}, the ergodic rate region of the MIMO-MAC is studied. In this section, we introduce our third main result: an iterative water-filling algorithm to describe the boundary of the ergodic rate region of the MIMO-MAC. In Section \ref{sec:simu}, we provide simulation results of the previously derived theoretical formulas. Finally, in Section \ref{sec:conclusion}, we give our conclusions.

{\it Notation:} In the following, boldface lower-case characters represent vectors, capital boldface characters denote matrices (${\I}_N$ is the $N\times N$ identity matrix). $X_{ij}$ denotes the $(i,j)$ entry of $\X$. The Hermitian transpose is denoted $(\cdot)^{\herm}$. The operators $\tr\X$, $|\X|$ and $\Vert \X\Vert$ represent the trace, determinant and spectral norm of matrix ${\X}$, respectively. The symbol $\EE\cdot]$ denotes expectation. The notation $F^{\Y}$ stands for the (cumulative) distribution function of the eigenvalues of the Hermitian matrix $\Y$. The function $(x)^+$ equals $\max(x,0)$ for real $x$. For $F$, $G$ two distribution functions, we denote $F\Rightarrow G$ the vague convergence of $F$ to $G$. The notation $x_n\asto x$ denotes the almost sure convergence of the sequence $x_n$ to $x$. The notation $\|F\|$ for the distribution function $F$ is the supremum norm defined as $\|F\|=\sup_x F(x)$. The symbol $\X\geq 0$ for a square matrix $\X$ means that $\X$ is Hermitian nonnegative definite.

\section{Scope and Summary of Main Results}
\label{sec:summary}
In this section, we summarise the main results of this article and explain their impact on the study of the effects of channel correlation on the achievable communication rates in the present multi-user framework.

\subsection{General Model}
\begin{figure}
  \centering
  \includegraphics[width=12cm]{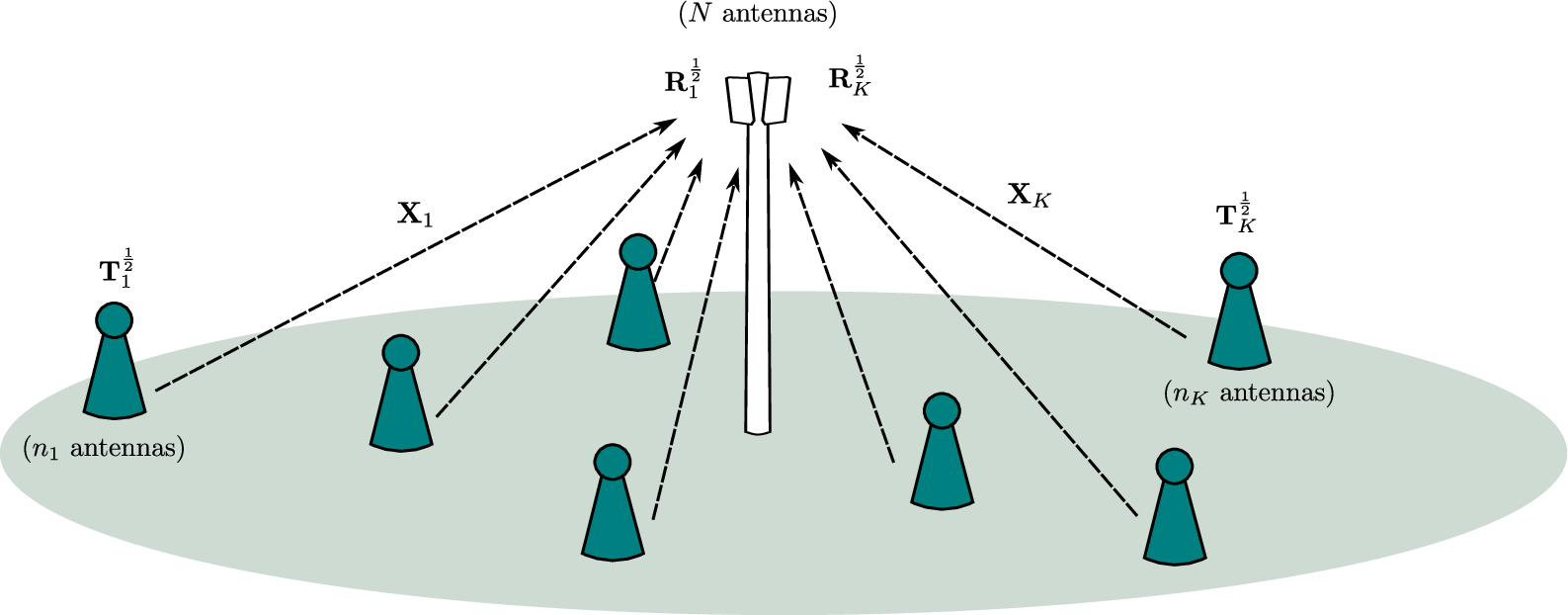}
  \caption{Multi-antenna multiple access scenario with Kronecker channels.}
  \label{fig:capaBC}
\end{figure}

Consider a set of $K$ wireless terminals, equipped with $n_1,\ldots,n_K$ antennas respectively, which we refer to as the transmitters, and another device equipped with $N$ antennas, which we call the receiver or the base station. We consider the uplink communication from the terminals to the base station. Denote $\H_k\in\CC^{N\times n_k}$ the channel matrix between transmitter $k$ and the receiver. Let $\H_k$ be defined as
\begin{equation}
\H_k = \R_k^\oh\X_k\T_k^\oh,
\end{equation}
where $\R_k^\oh\in\CC^{N\times N}$ and $\T_k^\oh\in\CC^{n_k\times n_k}$ are nonnegative Hermitian matrices and $\X_k\in\CC^{N\times n_k}$ is a realisation of a random matrix with independent Gaussian entries of zero mean and variance $1/n_k$. In this scenario, the matrices $\T_k$ and $\R_k$ model the correlation present in the channel at transmitter $k$ and at the receiver, respectively. This setup is depicted in Figure \ref{fig:capaBC}.

It is important to underline that the correlation patterns emerge both from the inter-antenna spacings on the volume-limited transmit and receive radio devices and from the solid angles of transmitted and received signal energy. Even though the transmit antennas emit signals in an isotropic manner, only a limited solid angle of emission is effectively received, and the same holds for the receiver which captures signal energy in a non-isotropic manner. Given this propagation factor, it is clear that the transmit covariance matrices $\R_k$ matrices are not equal for all users. We nonetheless assume physically identical and interchangeable antennas on each device. We therefore claim that the diagonal entries of $\R_k$ and $\T_k$, i.e., the variance of the channel fading on every antenna, are identical and equal to one, which, along with the normalisation of the Gaussian matrix $\X_k$, allows for a consistent definition of the SNR. As a consequence, $\tr \R_k=N$ and $\tr \T_k=n_k$. We will see that under these trace constraints the hypotheses made in Theorem \ref{th:1} are always satisfied, therefore making Theorem \ref{th:1} valid for all possible figures of correlation, including strongly correlated patterns. The hypotheses of Theorem \ref{th:2}, used to characterise the ergodic rate region of the MIMO-MAC, require additional mild assumptions, making Theorem \ref{th:2} valid for most practical models of $\R_k$ and $\T_k$. These statements are of major importance and rather new since in other contributions, e.g., \cite{HAC07}, \cite{DUP10}, it is usually assumed that the correlation matrices have uniformly bounded spectral norms across $N$. Physically this means that only low correlation patterns are allowed, excluding short distances between antennas and small solid angles of energy propagation. The counterpart of this interesting property is a theoretical reduction of the convergence rates of the derived deterministic equivalents, compared to those proposed in \cite{HAC07} and \cite{DUP10}. 

The rate performance of multi-cell or multi-user communication schemes is connected to the so-called Stieltjes transform and Shannon transform of matrices $\B_N$ of the type
\begin{equation}
  \label{eq:BN}
  \B_N = \sum_{k=1}^K \R_k^\oh\X_k\T_k\X_k^\herm\R_k^\oh.
\end{equation}
We study these matrices using tools from the field of large dimensional random matrix theory \cite{SIL06}. Among these tools, we define the Stieltjes transform $m_N(z)$ of the Hermitian nonnegative definite matrix $\B_N\in\CC^{N\times N}$, for $z\in \CC\setminus\RR^+$, as
\begin{align*}
  m_N(z) &= \int \frac1{\lambda - z}dF_N(\lambda) = \frac1N\tr\left(\B_N - z\I_N \right)^{-1},
\end{align*}
where $F_N$ denotes the (cumulative) distribution function of the eigenvalues of $\B_N$. The Stieltjes transform was originally used to characterise the asymptotic distribution of the eigenvalues of large dimensional random matrices \cite{MAR67}. From a wireless communications viewpoint, it can be used to characterise the signal-to-interference plus noise ratio of certain communication models, e.g., \cite{TSE99}, \cite{DEB03}. A second interest of the Stieltjes transform in wireless communications is its link to the so-called Shannon transform ${\mathcal V}_N(x)$ of $\B_N$, that we define for $x>0$ as
\begin{align*}
  {\mathcal V}_N(x) &= \frac1N\log\det\left(\I_N + \frac1x \B_N\right) = \int_0^{+\infty} \log\left(1+\frac{\lambda}x\right)dF_N(\lambda) = \int_x^{+\infty} \left(\frac1w - m_N(-w) \right)dw.
\end{align*}
The Shannon transform is commonly used to provide approximations of capacity expressions in large dimensional systems, e.g., \cite{TUL05}. In the present work, the Shannon transform of $\B_N$ will be used to provide a deterministic approximation of the ergodic achievable rate of the MIMO-MAC.

Before introducing our main results, namely Theorem \ref{th:1} and Theorem \ref{th:2}, which are rather technical and difficult to fathom without a preliminary explanation, we briefly describe these results in telecommunication terms and their consequences to the multi-user multi-cell communication models at hand.

\subsection{Main results}
The main results of this work unfold as follows:
\begin{itemize}
  \item Theorem \ref{th:1} provides a deterministic equivalent $m_N^\circ(z)$ for the Stieltjes transform $m_N(z)$ of $\B_N$, under the assumption that $N$ and $n_k$ grow large with the same order of magnitude and the sequences of distribution functions $\{F^{\T_k}\}_{n_k}$ and $\{F^{\R_k}\}_N$ form {\it tight sequences} \cite{BIL08}. This is, we provide an approximation $m_N^\circ(z)$ of $m_N(z)$ which can be expressed without reference to the random $\X_k$ matrices and which is almost surely asymptotically exact when $N\to \infty$. The tightness hypothesis is the key assumption that allows degenerated $\R_k$ and $\T_k$ matrices to be valid in our framework, and that therefore allows us to study strongly correlated channel models;
  \item Theorem \ref{th:2} provides a deterministic equivalent $\mathcal V_N^\circ(x)$ for the Shannon transform ${\mathcal V}_N(x)$ of $\B_N$. In this theorem, the assumptions on the $\R_k$ and $\T_k$ matrices are only slightly more constraining and of marginal importance for practical purposes. In particular, Theorem \ref{th:2} theoretically allows the largest eigenvalues of $\T_k$ or $\R_k$ to grow linearly with $N$, as the number of antennas increases, as long as the number of these large eigenvalues is of order $o(N)$;
  \item based on Theorem \ref{th:2}, the precoders that maximise the deterministic equivalent of the ergodic sum rate of the MIMO-MAC are computed. Those precoders have the following properties:
\begin{itemize}
  \item the eigenspace of the precoder for user $k$ coincides with the eigenspace of the transmit channel correlation matrix at user $k$;
  \item the eigenvalues of the precoder for user $k$ are the solution of a water-filling algorithm;
  \item as the system dimensions grow large, the mutual information achieved using these precoders becomes asymptotically close to the channel capacity.
\end{itemize}
\end{itemize}

The major practical interest of Theorems \ref{th:1} and \ref{th:2} lies in the possibility to analyze mutual information expressions for multi-dimensional channels, not as the averaged of stochastic variables depending on the random matrices $\X_k$ but as approximated deterministic expressions which do no longer feature the matrices $\X_k$. The study of those quantities is in general simpler than the study of the averaged stochastic expressions, which leads here to a simple derivation of the approximated rate optimal precoders.

In the next section, we introduce our theoretical results, whose proofs are left to the appendices.

\section{Mathematical Preliminaries}
\label{sec:maths}
In this section, we first introduce Theorem \ref{th:1}, which provides a deterministic equivalent for the Stieltjes transform of matrices $\B_N$ defined in \eqref{eq:BN}. A deterministic equivalent for the Shannon transform of $\B_N$ is then provided in Theorem \ref{th:2}, before we discuss in detail how this last result can be used to characterise the performance of the MIMO-MAC with strong channel correlation patterns.

\subsection{Main results}

\begin{theorem}
  \label{th:1}
  Let $K,N,n_1,\ldots,n_K$ be positive integers and let 
  \begin{equation}
    \B_N=\sum_{k=1}^K \R_k^\oh\X_k\T_k\X_k^{\herm}\R_k^\oh+\S
  \end{equation}
  be an $N\times N$ matrix with the following hypothesis for all $k\in \{1,\ldots,K\}$:
  \begin{enumerate}
	  \item \label{item:X} $\X_k\in\CC^{N\times n_k}$ has i.i.d. entries $\frac{1}{\sqrt{n_k}}X^k_{ij}$, such that $\EE[|X^k_{11}-\EE X^k_{11}|^2]=1$;
	  \item \label{item:R} $\R_k^\oh\in\CC^{N\times N}$ is a Hermitian nonnegative square root of the nonnegative definite Hermitian matrix $\R_k$;
    \item \label{item:T} $\T_k=\diag(\tau_1,\ldots,\tau_{n_k})$ with $\tau_i\geq 0$ for all $i$;
    \item \label{item:tight} the sequences $\{F^{\T_k}\}_{n_k}$ and $\{F^{\R_k}\}_{N}$ are tight;\footnote{this is, for all $\varepsilon>0$, there exists $M>0$ such that $F^{\T_k}(M)>1-\varepsilon$ and $F^{\R_k}(M)>1-\varepsilon$ for all $n_k$, $N$. See e.g., \cite{BIL08} for more details.}
    \item \label{item:S} $\S\in\CC^{N\times N}$ is Hermitian nonnegative definite;
    \item \label{item:c} there exist $b>a>0$ for which 
      \begin{equation}
	a< \min_k\lim\inf_N c_k\leq \max_k\lim\sup_N c_k< b,
      \end{equation}
      with $c_k=N/n_k$.
  \end{enumerate}
  Also denote, for $z\in \CC\setminus\RR^+$, $m_N(z)=\int \frac1{\lambda-z}dF_N(\lambda)$, the Stieltjes transform of $\B_N$. Then, as all $n_k$ and $N$ grow large, with ratio $c_k$,
  \begin{equation}
    \label{eq:th1conv}
    m_N(z) - m_N^\circ(z)\asto 0,
  \end{equation}
where
  \begin{align*} 
     m_N^\circ(z) = \frac{1}{N}\tr \left(\S + \sum_{k=1}^K\int\frac{\tau_k dF^{\T_k}(\tau_k)}{1+c_k\tau_k e_k(z)}\R_k -z\I_N \right)^{-1}
  \end{align*}
and the functions $e_i(z)$, $i\in \{1,\ldots,K\}$, form the unique solution to the $K$ equations
  \begin{align} 
    \label{eq:th1en}  e_i(z)=\frac{1}{N}\tr \R_i\left(\S + \sum_{k=1}^K\int\frac{\tau_k dF^{\T_k}(\tau_k)}{1+c_k\tau_k e_k(z)}\R_k -z\I_N \right)^{-1}
  \end{align}
  such that $\sgn(\Im [e_i(z)])=\sgn(\Im [z])$ when $\Im[z]\neq 0$ and such that $e_i(z)>0$ when $z$ is real and negative.

  Moreover, for any $\varepsilon>0$, the convergence of Equation \eqref{eq:th1conv} is uniform over any region of $\CC$ bounded by a contour interior to
  \begin{equation}
    \CC\setminus \left(\left\{z:|z|\leq \varepsilon \right\} \cup \left\{z=x+iv:x>0, |v|\leq\varepsilon \right\} \right).
  \end{equation}

  For all $N$, the function $m_N^\circ$ is the Stieltjes transform of a distribution function $F_N^\circ$. Denoting $F_N$ the empirical eigenvalue distribution function of $\B_N$, we finally have
  \begin{equation}
    F_N-F_N^\circ \Rightarrow 0
  \end{equation}
weakly and almost surely as $N\to \infty$.
\end{theorem}
\begin{IEEEproof}
The proof of Theorem \ref{th:1} is deferred to Appendix \ref{ap:th1}.
\end{IEEEproof}

A few technical remarks are of order at his point.
\begin{remark}
  In her PhD dissertation \cite{ZHA09}, Zhang derives an expression of the limiting eigenvalue distribution for the simpler case where $K=1$ and $\S=0$ but $\T_1$ is not constrained to be diagonal. Her work also uses a method based on the Stieltjes transform. Based on \cite{ZHA09}, it seems to the authors that Theorem \ref{th:1} could well be extended to non-diagonal $\T_k$. However, proving so requires involved calculus, which we did not perform. 
  Also, in \cite{SIL95}, using the same techniques as in the proof provided in Appendix \ref{ap:th1}, Silverstein {\it et al.} do not assume that the matrices $\T_k$ are nonnegative definite. Our result could be extended to this less stringent requirement on the central $\T_k$ matrices, although in this case Theorem \ref{th:1} does not hold for $z$ real negative. For application purposes though, it is fundamental that the Stieltjes transform of $\B_N$ exist for $z\in \RR^-$.
\end{remark}

We now claim that, under proper initialisation, for $z\in \CC\setminus \RR^+$, a classical fixed-point algorithm converges surely to the solution of \eqref{eq:th1en}. 
\begin{proposition}
  \label{prop:convergence}
  For $z\in \CC\setminus \RR^+$, the output $\{e^n_1,\ldots,e^n_K\}$ of fixed-point algorithm described in Table \ref{algo:fp} converges surely to the unique solution $\{e_1(z),\ldots,e_K(z)\}$ of \eqref{eq:th1en}, such that $\sgn(\Im [e_i(z)])=\sgn(\Im [z])$ when $\Im[z]\neq 0$ and such that $e_i(z)>0$ when $z<0$, for all $i$.
\end{proposition}
\begin{IEEEproof}
  The proof of Proposition \ref{prop:convergence}, inspired by the work of Dupuy {\it et al.} \cite{DUP10} in the context of frequency selective channel models, is provided in Appendix \ref{ap:fp}.  
\end{IEEEproof}

\begin{table}[t]
  \centering
  \begin{tabular}{l}
    \toprule
    \begin{minipage}[h]{0.6\linewidth}
      \begin{algorithmic}
	      \STATE Define $\varepsilon>0$, the convergence threshold and $n\geq 0$, the iteration step. For all $k\in \{1,\ldots,K\}$, set $e_j^0 =-1/z$ and $e_j^{-1}=\infty$.
	\WHILE{$\max_j \{|e_j^n-e_j^{n-1}|\}>\varepsilon$}
	\FOR{$k\in \{1,\ldots,K\}$}
	\STATE Compute
	\begin{equation}
		e^{n+1}_k = \frac{1}{N}\tr \R_k\left(\S + \sum_{j=1}^K\int\frac{\tau_j dF^{\T_j}(\tau_j)}{1+c_j\tau_j e^n_j}\R_j - z\I_N \right)^{-1} 
	\end{equation}
	\ENDFOR
	\STATE Assign $n\leftarrow n+1$
	\ENDWHILE
      \end{algorithmic}
    \end{minipage}\\
    \bottomrule
  \end{tabular}
  \caption{Fixed-point algorithm converging to the solution of \eqref{eq:th1en}}
  \label{algo:fp}
\end{table}


Different hypotheses will be used in the applications of Theorem \ref{th:1} provided in Section \ref{sec:capa}. For practical reasons, we will in particular need the entries of $\X_k$ will be Gaussian, the matrices $\T_k$ to be non-diagonal and $\S=0$. This entails the following corollary:
\begin{corollary}
  \label{co:1}
  Let $K,N,n_1,\ldots,n_K$ be positive integers and let
  \begin{equation}
    \B_N=\sum_{k=1}^K \R_k^\oh\X_k\T_k\X_k^{\herm}\R_k^\oh
  \end{equation}
  be an $N\times N$ matrix with the following hypothesis for all $k\in \{1,\ldots,K\}$:
  \begin{enumerate}
    \item $\X_k\in \CC^{N\times n_k}$ has i.i.d. Gaussian entries $\frac{1}{\sqrt{n_k}}X^k_{ij}$, with $\EE [X^k_{11}]=0$ and $\EE [|X^k_{11}|^2]=1$;
	  \item $\R_k^\oh\in\CC^{N\times N}$ is a Hermitian nonnegative square root of the nonnegative definite Hermitian matrix $\R_k$;
	  \item $\T_k\in\CC^{n_k\times n_k}$ is a nonnegative definite Hermitian matrix;
    \item \label{item:cotight} $\{F^{\T_k}\}_{n_k}$ and $\{F^{\R_k}\}_{N}$ form tight sequences;
    \item \label{item:ba0} there exist $b>a>0$ for which 
      \begin{equation}
	a< \min_k\lim\inf_N c_k\leq \max_k\lim\sup_N c_k< b
      \end{equation}
      with $c_k=N/n_k$.
  \end{enumerate}
Also denote, for $x>0$, $m_N(-x)=\frac{1}{N}\tr(\B_N+x\I_N)^{-1}$. Then, as all $N$ and $n_k$ grow large (while $K$ is fixed) 
  \begin{equation*}
    m_N(-x) - m_N^\circ(-x) \asto 0,
  \end{equation*}
where
  \begin{align*}
     m_N^\circ(-x) = \frac{1}{N}\tr \left(x\left[\I_N + \sum_{k=1}^K \delta_k(-x)\R_k \right]\right)^{-1}
  \end{align*}
  and the set of functions $\{e_i(-x),\delta_i(-x)\}$, $i\in \{1,\ldots,K\}$, form the unique solution to the equations
  \begin{align*} 
    e_i(-x)&=\frac{1}{N}\tr \R_i\left(x\left[\I_N + \sum_{k=1}^K \delta_k(-x)\R_k \right]\right)^{-1} \nonumber \\
    \delta_i(-x) &= \frac1{n_i}\tr \T_i\left( x\left[\I_{n_i} + c_ie_i(-x)\T_i \right]\right)^{-1},
  \end{align*}
  such that $e_i(-x)>0$ for all $i$.
\end{corollary}
\begin{IEEEproof}
  Since the $\X_k$ are Gaussian, the joint distribution of the entries of $\X_k\U$ coincides with that of $\X_k$, for $\U$ any $n_k\times n_k$ unitary matrix. Therefore, $\X_k\T_k\X_k^\herm$ in Theorem \ref{th:1} can be substituted by $\X_k(\U\T_k\U^\herm)\X_k^\herm$ without compromising the final result. As a consequence, the $\T_k$ can be taken nonnegative definite Hermitian and the result of Theorem \ref{th:1} holds. It then suffices to replace $\delta_i(-x)$ in the expression of $e_i(-x)$ to fall back on the result of Theorem \ref{th:1}.
\end{IEEEproof}

The deterministic equivalent of the Stieltjes transform $m_N$ of $\B_N$ is then extended to a deterministic equivalent of the Shannon transform of $\B_N$ in the following result:
\begin{theorem}
  \label{th:2}
   Let $x>0$ and $\B_N$ be a random Hermitian matrix as defined in Corollary \ref{co:1} with the following additional assumptions:
  \begin{enumerate}
    \item there exists $\alpha>0$ and a sequence $r_1,r_2,\ldots$, such that, for all $N$, 
      \begin{equation*}
	\max_{1\leq k\leq K} \max (\lambda^{\T_k}_{r_N+1},\lambda^{\R_k}_{r_N+1}) \leq \alpha,
      \end{equation*}
      where $\lambda^{\Z}_1\geq \ldots \geq \lambda^{\Z}_N$ denote the ordered eigenvalues of the $N\times N$ matrix $\Z$.
    \item \label{item2:tight} denoting $b_N$ an upper-bound on the spectral norm of the $\T_k$ and $\R_k$, $k\in \{1,\ldots,K\}$, and $\beta>0$ a constant such that $\beta>\frac{Kb}{a}(1+\sqrt{a})^2$, $a_N=b_N^2\beta$ satisfies 
      \begin{equation*}
	r_N\log(1+a_N/x)=o(N).
      \end{equation*}
  \end{enumerate}
Then, for large $N$, $n_k$, the Shannon transform ${\mathcal V}_N(x)=\int \log(1+\frac1x\lambda)dF_N(\lambda)$ of $\B_N$, satisfies
\begin{equation*}
	{\mathcal V}_N(x)-\mathcal V_N^\circ(x) \asto 0;
\end{equation*}
where
\begin{align}
    \label{eq:th2V2}
    \mathcal V_N^\circ(x) &= \sum_{k=1}^K \frac1N\log\det\left(\I_{n_k} + c_ke_k(-x)\T_k\right) + \frac1N\log\det\left(\I_N + \sum_{k=1}^K \delta_k(-x)\R_k\right)  - x \sum_{k=1}^K \delta_k(-x)e_k(-x).
  \end{align}
\end{theorem}
\begin{IEEEproof}
The proof of Theorem \ref{th:2} is provided in Appendix \ref{ap:th2}.
\end{IEEEproof}

Note that this last result is consistent both with \cite{TUL05} when the transmission channels are i.i.d. Gaussian and with \cite{MOU03} when $K=2$. This result is also similar in nature to the expressions obtained in \cite{HAC07} for the multi-antenna Rician channel model and with \cite{MOU07} in the case of frequency selective channels. We point out that the expressions obtained in \cite{MOU07}, \cite{DUP10} and \cite{DUM10}, when the entries of the $\X_k$ matrices are Gaussian distributed, suggest a faster convergence rate of the deterministic equivalent of the Stieltjes and Shannon transforms than the one obtained here. Indeed, while we show here a convergence of order $o(1)$ (which is in fact refined to $o(1/\log^kN)$ for any $k$ in Appendix \ref{ap:th1}), in those works the convergence is proved to be of order $O(1/N^2)$.

However, contrary to the above contributions, we allow the $\R_k$ and $\T_k$ matrices to be more general than uniformly bounded in spectral norm. This is thoroughly discussed in the section below.

\subsection{Kronecker channel with strong correlation patterns}
Theorem \ref{th:1} and Corollary \ref{co:1} require $\{F^{\R_k}\}_{N}$ and $\{F^{\T_k}\}_{n_k}$ to form tight sequences. Remark that, because of the trace constraint $\frac1N\tr \R_k = 1$, all sequences $\{F^{\R_k}\}_{N}$ are necessarily tight (the same reasoning naturally holds for $\T_k$). Indeed, given $\varepsilon>0$, take $M=2/\varepsilon$; $N[1-F^{\R_k}(M)]$ is the number of eigenvalues in $\R_k$ larger than $2/\varepsilon$, which is necessarily less than or equal to $N\varepsilon/2$ from the trace constraint, leading to $1-F^{\R_k}(M)\leq \varepsilon/2$ and then $F^{\R_k}(M)\geq 1-\varepsilon/2> 1 -\varepsilon$. The same naturally holds for the $\T_k$ matrices. Observe now that Condition \ref{item2:tight}) in Theorem \ref{th:2} requires a stronger assumption on the correlation matrices. Under the trace constraint, a sufficient assumption for Condition 2) is that there exists $\alpha>0$, such that the number of eigenvalues in $\R_k$ greater than $\alpha$ is of order $o(N/\log N)$. This is a mild assumption, which may not be verified for some very specific choices of $\{F^{\R_k}\}_{N}$.\footnote{As a counter-example, take $N=2^p+1$ and the eigenvalues of $\R_k$ to be 
\begin{equation*}
  2^{p-1},\underbrace{p,\ldots,p}_{\frac{2^{p-1}}p},\underbrace{0,\ldots,0}_{2^p-\frac{2^{p-1}}p}.
\end{equation*}
The largest eigenvalue is of order $N$ so that $a_N$ is of order $N^2$, and the number $r_N$ of eigenvalues larger than any $\alpha>0$ for $N$ large is of order $\frac{2^{p-1}}p\sim \frac{N}{\log(N)}$. Therefore $r_N\log(1+a_N/x)=O(N)$ here.}
Nonetheless, most conventional models for the $\R_k$ and $\T_k$, even when showing strong correlation properties, satisfy the assumptions of Theorem \ref{th:2}. We mention in particular the following examples:
\begin{itemize}
  \item if all $\R_k$ and $\T_k$ have uniformly bounded spectral norm, then there exists $\alpha>0$ such that all eigenvalues of $\R_k$ and $\T_k$ are less then $\alpha$ for all $N$. This implies $r_N=0$ for all $N$ and therefore the condition is trivially satisfied. Our model is therefore compatible with loosely correlated antenna structures;
  \item in contrast, when antennas are densely packed on a volume-limited device, the correlation matrices $\R_k$ and $\T_k$ tend to be asymptotically of finite rank, see e.g. \cite{POL03} in the case of a dense circular array. That is, for any given $\alpha>0$, for all large $N$, the number $r_N$ of eigenvalues greater than $\alpha$ is finite, while $a_N$ defined in Theorem \ref{th:2} is of order $N^2$. This implies $r_N\log(1+a_N/x)=O(\log N)=o(N)$ and therefore volume-limited devices with densely packed antennas are consistent with our framework;
  \item for one, two or three dimensional antenna arrays with neighbors separated by half the wavelength as discussed by Moustakas {\it et al.} in \cite{MOU00}, the correlation figures have a peculiar behaviour. In a linear array of antenna, $O(N)$ eigenvalues are of order of magnitude $O(1)$, the remaining eigenvalues being small. In a two-dimensional grid of antennas, $O(\sqrt{N})$ eigenvalues are of order $O(\sqrt{N})$, the remaining eigenvalues being close to zero. Finally, in a three-dimensional parallelepiped of antennas, $O(N^{\frac23})$ eigenvalues are of order $O(N^{\frac13})$, the remaining eigenvalues being close to $0$ also. As such, in the $p$-dimensional scenario, we can approximate $r_N$ by $N^{\frac{p-1}p}$, $a_N$ by $N^{\frac2p}$ and we have
    \begin{equation*}
      r_N\log(1+a_N/x) \sim N^{\frac{p-1}p}\log N = o(N),
    \end{equation*}
    so that the multi-dimensional antenna arrays with close antennas separated by half the wavelength also satisfy the hypotheses of Theorem \ref{th:2}.
\end{itemize}
As a consequence, a wide scope of antenna correlation models enter our deterministic equivalent framework, which comes again at the price of a slower theoretical convergence of the difference $\mathcal V_N-\mathcal V_N^\circ$.


We now move to practical applications of the above results and more specifically to the determination of the ergodic rate region of the MIMO-MAC.

\section{Rate Region of the MIMO-MAC}
\label{sec:capa}
In this section, we successively apply Theorem \ref{th:2} to approximate the ergodic mutual information for all deterministic precoders, and then we determine the precoders that maximise this approximated mutual information. This gives an approximation of all points on the boundary of the MIMO-MAC rate region. We also introduce an iterative power allocation algorithm to obtain explicitly the optimal precoders. 

\subsection{Deterministic equivalent of the mutual information}
Consider the wireless multiple access channel as described in Section \ref{sec:summary} and depicted in Figure \ref{fig:capaBC}. We denote $c_k=N/n_k$ the ratio between the number of antennas at the receive base station and the number of transmit antennas of user $k$. Denote $\s_k\in\CC^{n_k}$ the Gaussian signal transmitted by user $k$, such that $\EE[\s_k]=0$ and $\EE[\s_k\s_k^\herm]=\P_k$, with $\frac1{n_k}\tr\P_k\leq P_k$ where $P_k$ is the total power of transmitter $k$, $\y\in\CC^N$ the signal received at the base station and $\n$ the additive white Gaussian noise of variance $\EE[\n\n^\herm]=\sigma^2\I_N$. We recall that the Kronecker channel between user $k$ and the base station is denoted $\H_k=\R_k^\oh\X_k\T_k^\oh$, with the entries of $\X_k\in\CC^{N\times n_k}$ Gaussian independent of zero mean and variance $1/n_k$ and $\R_k$, $\T_k$ deterministic. The received vector $\y$ is therefore given by
\begin{equation*}
	  \y = \sum_{k=1}^K\H_k\s_k + \n.
\end{equation*}

Suppose that the $\H_k$ channels are varying fast and that the transmitters in the MAC only have statistical channel state information about the $\H_k$ in the sense that user $k$ only knows the long term statistics $\R_1,\ldots,\R_K$ and $\T_1,\ldots,\T_K$. In this case, for a noise variance equal to $\sigma^2$, the per-antenna ergodic MIMO-MAC rate region $\C_\MAC$ is given by \cite{GOL03}
\begin{align*}
  \C_\MAC &= \bigcup_{\substack{\frac1{n_i}\tr(\P_i)\leq P_i\\ \P_i\geq 0\\ i=1,\ldots,K }}\left\{\{R_i,1\leq i\leq K\}: \sum_{i\in \mathcal S}R_i
  \leq \EE \mathcal V_N(\P_{i_1},\ldots,\P_{i_{|\mathcal S|}};\sigma^2),\forall \mathcal S\subset\{1,\ldots,K\} \right\},
\end{align*}
with the expectation taken over the joint random variable $(\X_1,\ldots,\X_K)$, $\mathcal S=\{i_1,\ldots,i_{|\mathcal S|}\}$, and where we introduced the notation 
\begin{equation*}
  \mathcal V_N(\P_{i_1},\ldots,\P_{i_{|\mathcal S|}};\sigma^2) \triangleq \frac1N\log\det\left(\I_N + \frac{1}{\sigma^2}\sum_{i\in \mathcal S}\H_i\P_i\H_i^\herm \right).
\end{equation*}

Now, assuming the $\T_k$, $\P_k$ and $\R_k$ satisfy the hypotheses of Theorem \ref{th:2}, we have 
\begin{equation*}
\mathcal V_N(\P_{i_1},\ldots,\P_{i_{|\mathcal S|}};x)-\mathcal V_N^\circ(\P_{i_1},\ldots,\P_{i_{|\mathcal S|}};x)\to 0,
\end{equation*}
as $N,n_{i_1},\ldots,n_{i_{|\mathcal S|}}$ grow large for some sequence $\{\mathcal V_N^\circ(\P_{i_1},\ldots,\P_{i_{|\mathcal S|}};x)\}_N$, on a subset of measure $1$ of the probability space $\Omega$ that engenders $(\X_{i_1},\ldots,\X_{i_{|\mathcal S|}})$. Integrating this expression over $\Omega$ therefore leads to 
\begin{equation*}
\EE \mathcal V_N^\circ(\P_{i_1},\ldots,\P_{i_{|\mathcal S|}};x)-\mathcal V_N^\circ(\P_{i_1},\ldots,\P_{i_{|\mathcal S|}};x)\to 0.
\end{equation*}
We can therefore apply Theorem \ref{th:2} to determine the ergodic rate region $\C_\MAC$ of the MIMO-MAC. We specifically have
\begin{align}
  \label{eq:logdetmac_ergodic}
&\EE \mathcal V_N(\P_{i_1},\ldots,\P_{i_{|\mathcal S|}};\sigma^2) - \nonumber \\ &\left[
\sum_{k\in\mathcal S} \frac1N\log\det\left(\I_{n_k} + c_ke_k(-\sigma^2)\T_k\P_k\right) + \frac1N\log\det\left(\I_N + \sum_{k\in\mathcal S} \delta_k(-\sigma^2)\R_k\right)  - \sigma^2 \sum_{k\in\mathcal S} \delta_k(-\sigma^2)e_k(-\sigma^2)\right] \to 0,
\end{align}
with $e_i(-\sigma^2)$ and $\delta_i(-\sigma^2)$ the unique positive solutions of
\begin{align} 
  \label{eq:eiPI}    e_i(-\sigma^2)&=\frac{1}{N}\tr \R_i\left(\sigma^2\left[\I_N + \sum_{k\in\mathcal S} \delta_k(-\sigma^2)\R_k \right]\right)^{-1} \\
    \delta_i(-\sigma^2) &= \frac1{n_i}\tr \T_i^\oh\P_i\T_i^\oh\left( \sigma^2\left[\I_{n_i} + c_ie_i(-\sigma^2) \T_i^\oh\P_i\T_i^\oh\right]\right)^{-1}. \nonumber
\end{align}
This provides a deterministic equivalent for all points in the MIMO-MAC rate region, i.e., for all $\P_1,\ldots,\P_K$ precoders. 

\subsection{Rate maximisation}
Now we wish to determine for which precoders the boundary of the MIMO-MAC rate region is reached. This requires here to determine the rate optimal precoding matrices $\P_{i_1},\ldots,\P_{i_{|\mathcal S|}}$, for all $\mathcal S\subset \{1,\ldots,K\}$. To this end, we first need the following result:
\begin{proposition}
  \label{prop:concavity}
  If at least one of the correlation matrices $\T_k$, $k\in \mathcal S$, is invertible, then $\mathcal V_N^\circ$ is a strictly concave function in $\P_{i_1},\ldots,\P_{i_{|\mathcal S|}}$. 
\end{proposition}
\begin{IEEEproof}
The proof of Proposition \ref{prop:concavity} is provided in Appendix \ref{app:concavity}.
\end{IEEEproof}

Without loss of generality, for any $k$, since the $\X_k$ matrices are standard Gaussian, and therefore of unitarily invariant joint distribution, $\T_k$ can be assumed diagonal. If $\T_k$ is not of full rank then it can be reduced into a matrix of smaller size, such that the resulting matrix is invertible, without changing the problem at hand. We therefore assume all $\T_k$ matrices to be of full rank from now on. From Proposition \ref{prop:concavity}, we then immediately prove that the $|\mathcal S|$-ary set of matrices $\{\P_{i_1}^\circ,\ldots,\P^\circ_{i_{|\mathcal S|}}\}$ which maximises the deterministic equivalent of the ergodic sum rate over the set $\mathcal S$ is unique. In a very similar way as in \cite{DUM10}, we then show that the matrices $\P_k^\circ$, $k\in \mathcal S$, have the following properties:
\begin{proposition}
  \label{prop:wf}
  For every $k\in \mathcal S$, denote $\T_k=\U_k\bar{\T}_k\U_k^\herm$ the spectral decomposition of $\T_k$ with $\U_k$ unitary and $\bar{\T}_k=\diag(t_{k,1},\ldots,t_{k,n_k})$. Then the precoders $\P_{i_1}^\circ,\ldots,\P_{i_{|\mathcal S|}}^\circ$ which maximise the right-hand side of \eqref{eq:logdetmac_ergodic} satisfy:
  \begin{enumerate}
    \item $\P_k^\circ=\U_k\bar{\P}_k^\circ\U_k^\herm$, with $\bar{\P}_k^\circ$ diagonal, i.e., the eigenspace of $\P_k^\circ$ is the same as the eigenspace of $\T_k$;
    \item denoting, for all $k$, $e_k^\circ=e_k(-\sigma^2)$ as in \eqref{eq:eiPI} for $\P_k=\P_k^\circ$, the $i^{th}$ diagonal entry $p^\circ_{k,i}$ of $\bar{\P}_k^\circ$ satisfies:
      \begin{equation}
	\label{eq:qki}
	p^\circ_{k,i} = \left(\mu_k - \frac1{c_k e_k^\circ t_{ki}} \right)^+,
      \end{equation}
      where the $\mu_k$ are evaluated such that $\frac1{n_k}\tr\P_k=P_k$.
  \end{enumerate}
  In Table \ref{algo:1}, we provide an iterative water-filling algorithm to obtain the $p^\circ_{ki}$.
\end{proposition}
\begin{IEEEproof}
	The proof of Proposition \ref{prop:wf} is provided in Appendix \ref{ap:wf}.
\end{IEEEproof}

\begin{table}[t]
  \centering
  \begin{tabular}{l}
    \toprule
    \begin{minipage}[h]{0.6\linewidth}
      \begin{algorithmic}
	\STATE Define $\eta>0$ the convergence threshold and $l\geq 0$ the iteration step. At step $l=0$, for all $k\in \mathcal S$, $i\in\{1,\ldots,n_k\}$, set $p^0_{k,i}=P_k$.
	\WHILE{$\max_{k,i}\{ |p^l_{k,i}-p^{l-1}_{k,i}|\}>\eta$}
	\STATE For $k\in \mathcal S$, define $e_k^{l+1}$ as the solution of \eqref{eq:th1en} for $z=-\sigma^2$ and $\P_k$ with eigenvalues $p^l_{k,1},\ldots,p^l_{k,n_k}$, obtained from the fixed-point algorithm of Table \ref{algo:fp}.
	\FOR{$k\in \mathcal S$}
	\FOR{$i=1\ldots,n_k$}
	\STATE Set $p^{l+1}_{k,i}=\left(\mu_k - \frac1{c_ke^{l+1}_kt_{ki}} \right)^+$, with $\mu_k$ such that $\frac1{n_k}\tr\P_k=P_k$.
	\ENDFOR
	\ENDFOR
	\STATE Assign $l\leftarrow l+1$
	\ENDWHILE
      \end{algorithmic}
    \end{minipage}\\
    \bottomrule
  \end{tabular}
      \caption{Iterative water-filling algorithm}
  \label{algo:1}
\end{table}

\begin{remark}
In \cite{DUM10}, it is proved that the convergence of this algorithm implies its convergence towards the correct limit. The line of reasoning in \cite{DUM10} can be directly adapted to the current situation so that, if the iterative water-filling algorithm of Table \ref{algo:1} converges, then $\P_{i_1},\ldots,\P_{i_{|\mathcal S|}}$ converge to the matrices $\P_{i_1}^\circ,\ldots,\P_{i_{|\mathcal S|}}^\circ$. However, similar to \cite{DUM10}, it is difficult to prove the sure convergence of the water-filling algorithm. Nonetheless, extensive simulations suggest that convergence is always attained.
\end{remark}

For the set $\mathcal S$ under consideration, denote now $\P_{i_1}^\star,\ldots,\P_{i_{|\mathcal S|}}^\star$ the true sum rate maximising precoders. Then, if $\P_{i_1}^\circ,\ldots,\P_{i_{|\mathcal S|}}^\circ$ and $\P_{i_1}^\star,\ldots,\P_{i_{|\mathcal S|}}^\star$ are such that Condition 1) of Theorem \ref{th:2} is satisfied with the sets $\{\T_1,\ldots,\T_K\}$ and $\{\R_1,\ldots,\R_K\}$ replaced by $\{\T_{i_1}\P^\circ_{i_1},\ldots,\T_{i_{|\mathcal S|}}\P^\circ_{i_{|\mathcal S|}}\}$ (or $\{\T_{i_1}\P^\star_{i_1},\ldots,\T_{i_{|\mathcal S|}}\P^\star_{i_{|\mathcal S|}}\}$) and $\{\R_{i_1},\ldots,\R_{i_{|\mathcal S|}}\}$, respectively, we have from Theorem \ref{th:2}
\begin{align*}
\mathcal V_N(\P_{i_1}^\star,\ldots,\P_{i_{|\mathcal S|}}^\star;\sigma^2) - \mathcal V_N(\P_{i_1}^\circ,\ldots,\P_{i_{|\mathcal S|}}^\circ;\sigma^2) &= \left(\mathcal V_N(\P_{i_1}^\star,\ldots,\P_{i_{|\mathcal S|}}^\star;\sigma^2) - \mathcal V_N^\circ(\P_{i_1}^\star,\ldots,\P_{i_{|\mathcal S|}}^\star;\sigma^2) \right) \nonumber \\
&+ \left(\mathcal V_N^\circ(\P_{i_1}^\star,\ldots,\P_{i_{|\mathcal S|}}^\star;\sigma^2) - \mathcal V_N^\circ(\P_{i_1}^\circ,\ldots,\P_{i_{|\mathcal S|}}^\circ;\sigma^2) \right) \nonumber \\
&+ \left( \mathcal V_N^\circ(\P_{i_1}^\circ,\ldots,\P_{i_{|\mathcal S|}}^\circ;\sigma^2) - \mathcal V_N(\P_{i_1}^\circ,\ldots,\P_{i_{|\mathcal S|}}^\circ;\sigma^2) \right),
\end{align*}
where both right-hand side differences of the type $\mathcal V_N-\mathcal V_N^\circ$ tend to zero, while the left-hand side term is positive by definition of $\P_k^\star$ and the remaining right-hand side term is negative by definition of the $\P_k^\circ$. This finally ensures that
\begin{equation*}
  \mathcal V_N(\P_{i_1}^\star,\ldots,\P_{i_{|\mathcal S|}}^\star;\sigma^2) - \mathcal V_N(\P_{i_1}^\circ,\ldots,\P_{i_{|\mathcal S|}}^\circ;\sigma^2) \to 0,
\end{equation*}
as $N,n_{i_1},\ldots,n_{i_{|\mathcal S|}}$ grow large with uniformly bounded ratios. Therefore, the mutual information obtained based on the precoders $\P_{i_1}^\circ,\ldots,\P_{i_{|\mathcal S|}}^\circ$ is asymptotically close to the capacity achieved with the ideal precoders $\P_{i_1}^\star,\ldots,\P_{i_{|\mathcal S|}}^\star$. Finally, if, for all sets $\mathcal S\subset \{1,\ldots,K\}$, the matrices $\T_k$, $\R_k$ and the resulting $\P_k^\star$, $\P_k^\circ$ satisfy the mild conditions of Theorem \ref{th:2}, then all points of the boundary of the MIMO-MAC rate region can be given a deterministic equivalent.

This concludes this application section. In the following section, we provide simulation results that confirm the accuracy of the deterministic equivalents as well as the validity of the hypotheses made on the $\T_k$ and $\R_k$ matrices.

\section{Simulations and Results}
\label{sec:simu}

\begin{figure} 
  \centering

\includegraphics[]{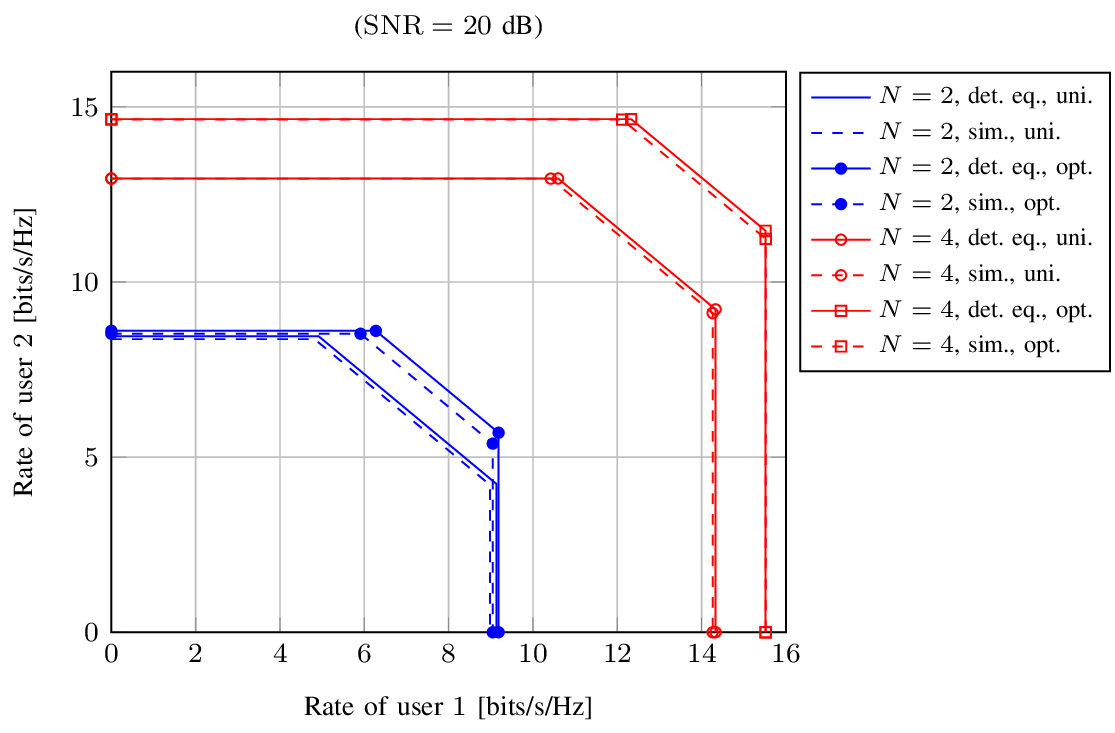}
\includegraphics[]{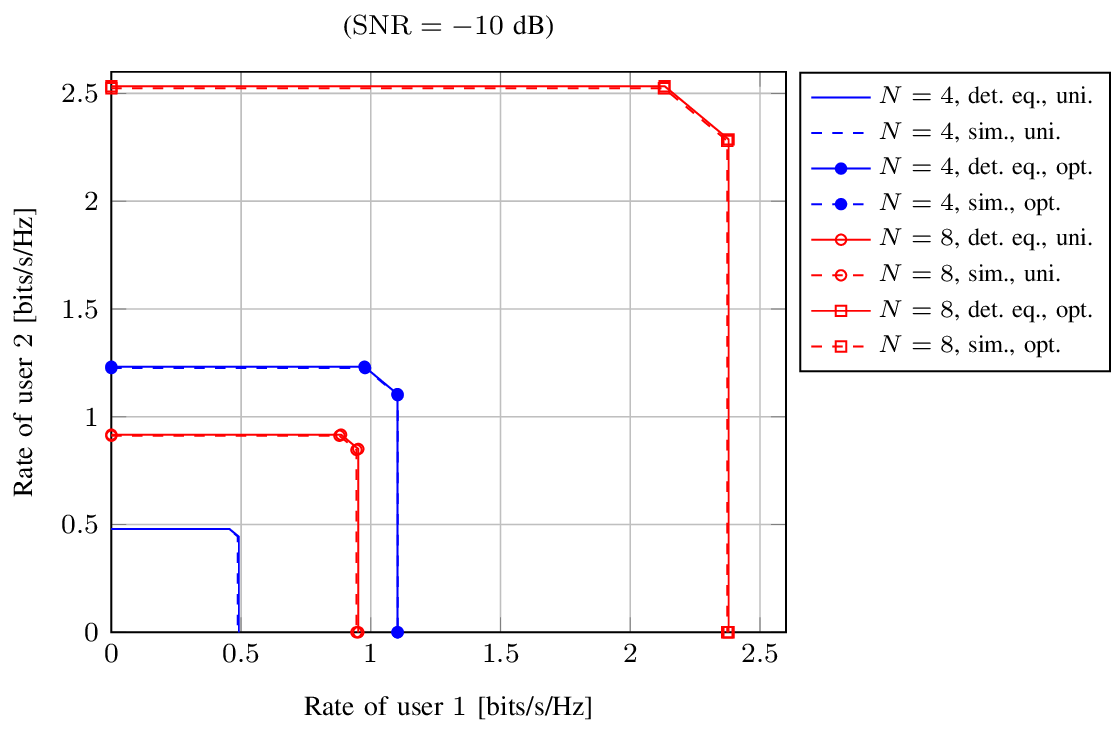}
    \caption{Rate region for a two-user MIMO-MAC when the antennas are placed on a linear array with distance between close antennas is $\lambda/10$. The number of antennas $N=n_1=n_2$ is taken equal to $2$, $4$ or $8$. Simulations (sim.) are compared against deterministic equivalents (det. eq.). Uniform power allocation across transmit antennas (uni.) as well as optimal sum rate maximising precoding (opt.) are considered. The SNR is $20$ dB in the top figure and $-10$ dB in the bottom figure. }
    \label{fig:ergodicrate}
  \end{figure}

\begin{figure} 
  \centering
\includegraphics[]{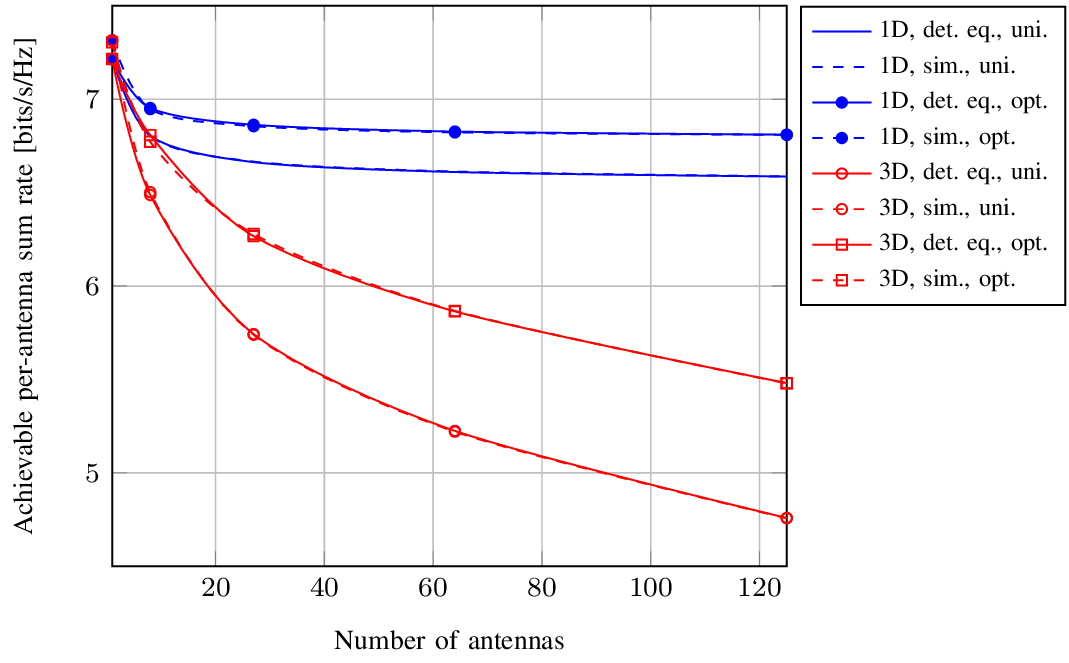}
    \caption{Per-antenna sum rate for a two-user MIMO-MAC when the antennas are placed on a line (1D) or a cubic array (3D). The number of antennas satisfy $N=n_1=n_2$ and range from $1$ to $125$. Simulations (sim.) are compared against deterministic equivalents (det. eq.). The distance between close antennas is $\lambda/2$. The SNR is $20$ dB. Uniform power allocation (uni.) as well as optimal sum rate maximising precoding (opt.) are considered.}
    \label{fig:antenna_efficiency}
  \end{figure}

  In the following, we apply the results obtained in Section \ref{sec:capa} to provide comparative simulation results between ergodic rate regions, sum rates and their respective deterministic equivalents, for non negligible channel correlations on both communication sides. We provide simulation results in the context of a two-user MIMO-MAC, with $N$ antennas at the base station and $n_1=n_2$ antennas at the user terminals. The antennas are placed on a possibly multi-dimensional array, antenna $i$ being located at $\x_i\in\RR^3$. We further assume that both terminals are physically identical. To model the transmit and receive correlation matrices, we consider both the effect of the distances between adjacent antennas at the user terminals and at the base station, and the effect of the solid angles of effective energy transmission and reception. We assume a channel model where signals are transmitted and received isotropically in the vertical direction, but transmitted and received under an angle $\pi$ in the horizontal direction. We then model the entries of the correlation matrices from a natural extension of Jakes model \cite{CLA68} with privileged direction of signal departure and arrival. Denoting $\lambda$ the transmit signal wavelength, $T_{1_{ab}}$, the entry $(a,b)$ of the matrix $\T_1$, is 
\begin{equation*}
  \label{eq:jakes}
  T_{1_{ab}}=\int_{\theta_{\min}^{(\T_1)}}^{\theta_{\max,}^{(\T_1)}}\exp\left(\frac{2\pi i}{\lambda} \Vert\x_a-\x_b\Vert \cos(\theta)\right)d\theta
\end{equation*}
with $[\theta_{\min}^{(\T_1)},\theta_{\max}^{(\T_1)}]$ the effective horizontal directions of signal propagation. With similar notations for the other correlation matrices, we choose $\theta_{\min}^{(\T_1)}=0$, $\theta_{\max}^{(\T_1)}=\pi$, $\theta_{\min}^{(\T_2)}=\pi/3$, $\theta_{\max}^{(\T_2)}=4\pi/3$, $\theta_{\min}^{(\R_1)}=2\pi/3$, $\theta_{\max}^{(\R_1)}=-2\pi/3$, $\theta_{\min}^{(\R_2)}=\pi$ and $\theta_{\max}^{(\R_2)}=0$.

We start by simulating the MIMO-MAC rate region obtained when $n_1=n_2=N$, either for $N=2$, $N=4$ or $N=8$, for a linear antenna array with distance $\lambda/10$ between close antennas, under identity or optimal precoding policies and with signal to noise ratio of $20$ dB or $-10$ dB. Simulation results, averaged over $10,000$ channel realisations are compared with the deterministic equivalents. This is depicted in Figure \ref{fig:ergodicrate}. The deterministic equivalents of the rate regions appear to approximate the true rate regions extremely well, even for very small system dimensions. The case $N=8$ shows in particular a perfect match. We note that increasing the number of antennas on both communication side provides a greater gain when using the optimal precoding policy. We also observe that, while increasing the number of antennas tends to reduce the individual per-antenna rate under uniform power allocation, using an optimal precoding policy significantly increases the per-antenna rate. This phenomenon is particularly accentuated in the low SNR regime. This confirms the observations made by Vishwanath {\it et al.} in \cite{VIS03b}, according to which the efficiency of individual antennas can grow as the correlation grows at low SNR. This is, for a given number of antennas, by increasing correlation and systematically applying optimal precoding, strong eigenmodes emerge over which data can be directed. This leads to higher rates than for an uncorrelated antenna array for which there are no such strong eigenmodes.

In order to test the robustness of the proposed deterministic equivalents to strongly correlated channel conditions, we then compare in Figure \ref{fig:antenna_efficiency} the MIMO-MAC ergodic sum rate to the associated deterministic equivalents when precoder optimisation is performed or not and when the antenna grids are either one-dimensional regular arrays or three-dimensional regular cubes. In all situations, the distance between neighboring antennas is half the wavelength and the signal to noise ratio is taken to be $20$ dB. We take the number of antennas on both sides to be successively $1$, $8$, $27$, $64$ and $125$. We first observe that the deterministic equivalents are extremely accurate in this scenario. We confirm also the behaviour of the antenna efficiency, which saturates for the one-dimensional array and decreases fast for the three-dimensional antenna array, similar to what was observed in \cite{MOU00} for the single-user case. In terms of performance, a large improvement of the achievable sum rate is observed, especially in the three-dimensional case, when the optimal precoding policy is applied. Using optimal precoding policies can therefore significantly reduce the negative impact of antenna correlation even in this high SNR regime.

\section{Conclusion}
\label{sec:conclusion}
In this article, we analyzed the performance of multi-antenna multi-user wireless communications and more particularly the multi-antenna multiple access channel, while taking into account the correlation effects due to close antennas and reduced solid angles of energy transmission and reception. The analytic approach is based on novel results, based on recent tools from the field of large dimensional random matrix theory. From these tools, we provide on the one hand a deterministic equivalent of the per-antenna mutual information of the MIMO-MAC for arbitrary precoders under possibly strongly correlated channel conditions and on the other hand an approximation of the rate maximising precoders along with an iterative water-filling algorithm to compute these precoders. In particular, while theoretical results prove the asymptotic accuracy of our model in the case of dense antenna packing or multi-dimensional antenna array on either communication side, simulations concur and suggest that the deterministic equivalents are moreover extremely accurate for very small system dimensions. These results can be used both from a practical side to easily derive optimal precoders and from a theoretical side to quantify the gains achieved by optimal power allocation policies in strongly correlated MIMO channels.

\newpage
\appendices

\section{Proof of Theorem \ref{th:1}}
\label{ap:th1}
  For ease of read, the proof will be divided into several sections.

  We first consider the case $K=1$, whose generalisation to $K\geq 1$ is given in Appendix \ref{sec:proof_K}. Therefore, in the coming sections, we drop the useless indexes.
\subsection{Truncation and centralisation}
We begin with the truncation and centralisation steps which will replace $\X$, $\R$ and $\T$ by matrices with bounded entries, more suitable for analysis; the difference of the Stieltjes transforms of the original and new $\B_N$ converging to zero. Since vague convergence of distribution functions is equivalent to the convergence of their Stieltjes transforms, it is sufficient to show the original and new empirical distribution functions of the eigenvalues approach each other almost surely in the space of subprobability measures on $\RR$ with respect to the topology which yields vague convergence.

Let $\widetilde{X}_{ij}=X_{ij}{\bf 1}_{\{ |X_{ij}|<\sqrt{N}\}}-\EE(X_{ij}{\bf 1}_{\{ |X_{ij}|<\sqrt{N}\}})$ and $\widetilde{\X}=\left(\frac1{\sqrt{n}}\widetilde{X}_{ij}\right)$. Then, from c), Lemma \ref{le:1} and a), Lemma \ref{le:3}, it follows exactly as in the initial truncation and centralisation steps in \cite{SIL95} and \cite{DOZ07} (which provide more details in their appendices), that
\begin{equation*}
\| F_N-F^{\S+\R^\oh\widetilde \X \T\widetilde \X^\herm \R^\oh}\|\asto 0,
\end{equation*}
as $N\to\infty$.

Let now $\overline{X}_{ij}=\widetilde{X}_{ij}\cdot {\bf 1}_{\{|X_{ij}|<\ln N\}}-\EE(\widetilde{X}_{ij}{\bf 1}I_{\{|X_{ij}|<\ln N\}})$ and $\overline{\X}=\left(\frac1{\sqrt n}\overline{X}_{ij}\right)$.
This is the final truncation and centralisation step, which will be practically handled the same way as in \cite{SIL95}, which some minor modifications, given presently.

For any Hermitian non-negative definite $r\times r$ matrix $\A$, let $\lambda^\A_i$
denote its $i$-th smallest eigenvalue of $\A$. With $\A=\U\diag(\lambda_1^\A,\ldots,\lambda_r^\A)\U^\herm$ its spectral decomposition, let for any $\alpha>0$
\begin{equation*}
  \A^{\alpha}= \U\diag(\lambda_1^\A {\bf 1}_{\{\lambda_r^\A\leq\alpha\}},\ldots,\lambda_r^\A{\bf 1}_{\{\lambda_1\leq\alpha\}})\U^\herm.
\end{equation*}
Then for any $N\times N$ matrix $\Q$, we get from 1) and 2), Lemma \ref{le:3},
\begin{align*}
\|F^{\S+\R^\oh \Q\T\Q^\herm\R^\oh}-F^{\S+\R^{\frac12\alpha}Q\T^{\alpha}\Q^\herm\R^{\frac12\alpha}}\|
& \leq\frac2N\rank(\R^\oh-\R^{\frac12\alpha})+\frac1N\rank(\T-\T^{\alpha}) \\
&=\frac2N\sum_{i=1}^N {\bf 1}_{\{\lambda_i^\R>\alpha\}}+\frac1N\sum_{i=1}^n {\bf 1}_{\{\lambda_i^\T>\alpha)\}} \\
&=2F^\R((\alpha,\infty))+\frac1{c_N}F^\T((\alpha,\infty)).
\end{align*}

Therefore, from the assumptions \ref{item:tight}) and \ref{item:c}) in Theorem \ref{th:1}, we have for any sequence $\{\alpha_N\}$ with $\alpha_N\to\infty$
\begin{equation}
  \label{eq:1}
\|  F^{\S+\R^\oh \Q\T\Q^\herm\R^\oh}-F^{\S+\R^{\frac12\alpha_N}\Q\T^{\alpha_N}\Q^\herm\R^{\frac12\alpha_N}}\| \to 0,
\end{equation}
as $N\to\infty$.

A metric $D$ on probability measures defined on $\RR$, which induces the topology of vague convergence, is introduced in \cite{SIL95} to handle the last truncation step.  The matrices studied in \cite{SIL95} are essentially $\B_N$ with $\R=\I_N$. Following the steps beginning at (3.4) in \cite{SIL95}, we see in our case that when $\alpha_N$ is chosen so that as $N\to\infty$, $\alpha_N\uparrow\infty$,
\begin{equation*}
  \alpha_N^8(\EE|X_{11}^2{\bf 1}_{\{X_{11}|\ge\ln N\}}+N^{-1})\to 0
\end{equation*}
and
\begin{equation*}
  \sum_{N=1}^{\infty}\frac{\alpha_N^{16}}{N^2}\left(\EE|X_{11}|^4{\bf 1}_{\{|X_{11}|<\sqrt N)\}}+1\right)<\infty.
\end{equation*}

We will get
\begin{equation}
  \label{eq:2}
  D(F^{\S+\R^{\frac12\alpha_N}\widetilde{\X}\T^{\alpha_N}\widetilde{\X}^\herm\R^{\frac12\alpha_N}},F^{\S+\R^{\frac12\alpha_N}\overline{\X}\T^{\alpha_N}\overline{\X}^\herm\R^{\frac12\alpha_N}})\asto 0
\end{equation}
as $N\to\infty$.  

Since $\EE|\overline{X}_{11}|^2\to 1$ as $N\to\infty$ we
can rescale and replace $\overline{\X}$ with $\overline{\X}/\sqrt\EE|\overline{X}_{11}|^2$, whose components are bounded by $k\ln N$ for some $k>2$. Let $\log N$ denote logarithm of $N$ with base $e^{1/k}$ (so that $k\ln N=\log N$). 
Therefore, from \eqref{eq:1} and \eqref{eq:2} we can assume that for each $N$ the $X_{ij}$ are i.i.d., $\EE X_{11}=0$, $\EE|X_{11}|^2=1$, and $|X_{ij}|\leq\log N$.  

Later on the proof will require a restricted growth rate on both $\Vert \R\Vert$ and $\Vert \T\Vert$.  
We see from \eqref{eq:1} that we can also assume 
\begin{equation}
  \label{eq:3}
\max(\|\R\|,\|\T\|)\leq\log N.
\end{equation}

\subsection{Deterministic approximation of $m_N(z)$}
Write $\X=[\x_1,\ldots,\x_n]$, $\x_i\in \CC^N$ and let $\y_j=(1/\sqrt{n})\R^\oh \x_j$. Then we can write
\begin{equation*}
\B_N=\S+\sum_{j=1}^n \tau_j \y_j \y_j^\herm.
\end{equation*}
We assume $z\in\CC^+$ and let $v=\Im[z]$. Define
\begin{equation*}
e_N=e_N(z)=(1/N)\tr \R(\B_N-z\I_N)^{-1}
\end{equation*}
and
\begin{equation*}
p_N=-\frac{1}{nz}\sum_{j=1}^n\frac{\tau_j}{1+c_N \tau_j e_N}=\int\frac{-\tau}{z(1+c_N \tau e_N)}dF^{\T}(\tau)
\end{equation*}
Write $\B_N = \O\Lambda \O^\herm$, $\Lambda=\diag(\lambda_1,\ldots,\lambda_N)$, its spectral decomposition. Let $\underline \R=\{\underline R_{ij}\}=\O^\herm\R\O$.
Then
\begin{equation*}
e_N=(1/N)\tr\underline \R(\Lambda-z\I_N)^{-1}=(1/N)\sum_{i=1}^N\frac{\underline R_{ii}}{\lambda_i-z}.
\end{equation*}
We therefore see that $e_N$ is the Stieltjes transform of a measure on the nonnegative reals with total mass $(1/N)\tr \R$. It follows that both $e_N(z)$ and $ze_N(z)$ map $\CC^+$ into $\CC^+$. This implies that $p_N(z)$ and $zp_N(z)$ map $\CC^+$ into $\CC^+$ and, as $z\to\infty$, $zp_N(z)\to -(1/n)\tr \T$. Therefore, from Lemma \ref{le:5}, we also have $p_N$ the Stieltjes transform of a measure on the nonnegative reals with total mass $(1/n)\tr \T$. From \eqref{eq:3}, it follows that 
\begin{equation}
  \label{eq:4}
|e_N|\leq v^{-1}\log N
\end{equation}
and 
\begin{equation}
  \label{eq:5}
\left|\int\frac{\tau}{(1+ c_N \tau e_N)}dF^{\T}(\tau)\right|=|zp_N(z)|
\leq |z|v^{-1}\log N.
\end{equation}
More generally, from Lemma \ref{le:5}, any function of the form
\begin{equation*}
\frac{-\tau}{z(1+m(z))},
\end{equation*}
where $\tau \geq 0$ and $m(z)$ is the Stieltjes transform of a finite measure on $\RR^+$, is the Stieltjes transform of a measure on the nonnegative reals with total mass $\tau$. It follows that
\begin{equation}
  \label{eq:6}
\left|\frac{\tau}{1+m(z)}\right|\leq \tau|z|v^{-1}.
\end{equation}
Fix now $z\in\CC^+$. Let $\B_{(j)}=\B_N - \tau_j \y_j \y_j^\herm$. Define $\D=-z\I_N+\S-zp_N(z)\R$. We write
\begin{equation*}
\B_N-z\I_N-\D = \sum_{j=1}^n \tau_j \y_j \y_j^\herm + z p_N \R.
\end{equation*}
Taking inverses and using Lemma \ref{le:4} we have
\begin{align*}
(\B_N-z\I_N)^{-1}-\D^{-1} &=\sum_{j=1}^n \tau_j\D^{-1}\y_j\y_j^\herm(\B_N-z\I_N)^{-1}+zp_N\D^{-1}\R(\B_N-z\I_N)^{-1} \\
&=\sum_{j=1}^n\tau_j\frac{\D^{-1}\y_j\y_j^\herm(\B_{(j)}-z\I_N)^{-1}}{1 + \tau_j\y_j^\herm(\B_{(j)}-z\I_N)^{-1}\y_j} + zp_N\D^{-1}\R(\B_N-z\I_N)^{-1}.
\end{align*}
Taking traces and dividing by $N$, we have
\begin{equation*}
\frac1N\tr \D^{-1} - m_N(z)=\frac1n\sum_{j=1}^n\tau_j d_j\equiv w_N^m,
\end{equation*}
where
\begin{equation*}
d_j=\frac{(1/N)\x_j^\herm\R^\oh(\B_{(j)}-z\I_N)^{-1}\D^{-1}\R^\oh \x_j}
{1+\tau_j\y_j^\herm(\B_{(j)}-z\I_N)^{-1}\y_j}-\frac{(1/N)\tr \R(\B_N-z\I_N)^{-1}\D^{-1}}{1+c_N\tau_je_N}.
\end{equation*}
Multiplying both sides of the above matrix identity by $\R$, and then taking traces and dividing by $N$, we find
\begin{equation*}
\frac1N\tr \D^{-1}\R - e_N(z) = \frac1n\sum_{j=1}^n \tau_j d_j^e \equiv w_N^e,
\end{equation*}
where
\begin{equation*}
d_j^e=\frac{(1/N)\x_j^\herm\R^\oh(\B_{(j)}-z\I_N)^{-1}\R\D^{-1}\R^\oh \x_j}
{1+\tau_j\y_j^\herm(\B_{(j)}-z\I_N)^{-1}\y_j}-\frac{(1/N)\tr \R(\B_N-z\I_N)^{-1}\R\D^{-1}}{1+c_N\tau_je_N}.
\end{equation*}

We then show that, for any $k>0$, almost surely
\begin{equation}
  \label{eq:9}
\lim_{N\to\infty}(\log^kN)w^m_N=0
\end{equation}
and
\begin{equation}
  \label{eq:9b}
\lim_{n\to\infty}(\log^kN)\,w^e_N=0.
\end{equation}
Notice that for each $j$, $\y_j^\herm(\B_{(j)}-z\I_N)^{-1}\y_j$ can be viewed as the Stieltjes transform of a measure on $\RR^+$. Therefore from \eqref{eq:6} we have
\begin{equation}
  \label{eq:7}
\left|\frac{1}{1+\tau_j\y_j^\herm(\B_{(j)}-z\I_N)^{-1}\y_j}\right|\leq\frac{|z|}{v}.
\end{equation}
For each $j$, let $e_{(j)}=e_{(j)}(z)=(1/N)\tr \R(\B_{(j)}-z\I_N)^{-1}$, and
\begin{equation*}
p_{(j)}=p_{(j)}(z) =\int\frac{-\tau}{z(1+c_N\tau e_{(j)})}dF^{\T}(\tau),
\end{equation*}
both being Stieltjes transforms of measures on $\RR^+$, along with the integrand for each $\tau$.

Using Lemma \ref{le:4}, Equations \eqref{eq:3} and \eqref{eq:6}, we have
\begin{equation}
  \label{eq:8}
|zp_N-zp_{(j)}| = |e_N-e_{(j)}|c_N\left|\int\frac{\tau^2}{(1+c_N\tau e_N)(1+c_N\tau e_{(j)})}dF^{\T}(\tau)\right|\leq\frac{c_N|z|^2\log^3N}{Nv^3}.
\end{equation}
Let $\D_{(j)}=-z\I_N+\S-zp_{(j)}(z)\R$. Notice that $(\B_N-z\I_N)^{-1}$ and $(\B_{(j)}-z\I_N)^{-1}$ are  bounded in spectral norm by $v^{-1}$ and, from Lemma \ref{le:6}, the same holds true for $\D^{-1}$ and $\D_{(j)}^{-1}$.  

In order to handle both $w_N^m$, $d_j$ and $w_N^e$, $d^e_j$ at the same time, we shall denote by $\E$ either $\T$ or $\I_N$, and $w_N$, $d_j$ for now will denote either the original $w_N^m$, $d_j$ or $w_N^e$, $d^e_j$. Write $d_j=d^1_j+d^2_j+d_j^3+d_j^4$, where
\begin{align*}
d_j^1 &=
\frac{(1/N)\x_j^\herm\R^\oh(\B_{(j)}-z\I_N)^{-1}\E\D^{-1}\R^\oh \x_j}{1+\tau_j\y_j^\herm(\B_{(j)}-z\I_N)^{-1}\y_j}
-\frac{(1/N)\x_j^\herm \R^\oh(\B_{(j)}-z\I_N)^{-1}\E\D_{(j)}^{-1}\R^\oh \x_j}{1+\tau_j\y_j^\herm(\B_{(j)}-z\I_N)^{-1}\y_j} \\
d_j^2 &=
\frac{(1/N)\x_j^\herm \R^\oh(\B_{(j)}-z\I_N)^{-1}\E\D_{(j)}^{-1}\R^\oh\x_j}{1+\tau_j\y_j^\herm(\B_{(j)}-z\I_N)^{-1}\y_j}  
 -\frac{(1/N)\tr \R(\B_{(j)}-z\I_N)^{-1}\E\D_{(j)}^{-1}}{1+\tau_j\y_j^\herm(\B_{(j)}-z\I_N)^{-1}\y_j} \\
d_j^3 &=
\frac{(1/N)\tr \R(\B_{(j)}-z\I_N)^{-1}\E\D_{(j)}^{-1}}{1+\tau_j\y_j^\herm(\B_{(j)}-z\I_N)^{-1}\y_j} 
-\frac{(1/N)\tr \R(\B_N-z\I_N)^{-1}\E\D^{-1}}{1+\tau_j\y_j^\herm(\B_{(j)}-z\I_N)^{-1}\y_j} \\
d_j^4 &=
\frac{(1/N)\tr \R(\B_N-z\I_N)^{-1}\E\D^{-1}}{1+\tau_j\y_j^\herm(\B_{(j)}-z\I_N)^{-1}\y_j}
 -\frac{(1/N)\tr \R(\B_N-z\I_N)^{-1}\E\D^{-1}}{1+c_N\tau_je_N}.
\end{align*}
From Lemma \ref{le:4}, Equations \eqref{eq:3}, \eqref{eq:7} and \eqref{eq:8}, we have
\begin{align*}
\tau_j|d_j^1| &\leq
\frac1N\|\x_j\|^2\frac{c_N\log^7N|z|^3}{Nv^7} \\
\tau_j|d_j^2| &\leq
|z|v^{-1}\frac{\log N}N\left|\x_j^\herm \R^\oh(\B_{(j)}-z\I_N)^{-1}\E\D_{(j)}^{-1}\R^\oh\x_j
 -\tr \R(\B_{(j)}-z\I_N)^{-1}\E\D_{(j)}^{-1}\right| \\
\tau_j|d_j^3| &\leq 
\frac{|z|\log^3N}{vN}\left(\frac{1}{v^2}+\frac{c_N|z|^2\log^3N}{v^6}\right)\to 0,\text{ as $n\to\infty$} \\
\tau_j|d_j^4| &\leq
\frac{|z|c_N\log^4N}{Nv^3}\left(|\x_j^\herm\R^\oh(\B_{(j)}-z\I_N)^{-1}\R^\oh \x_j
-\tr\R^\oh(\B_{(j)}-z\I_N)^{-1}\R^\oh|+\frac{\log N}v\right).
\end{align*}
From Lemma \ref{le:13}, there exists $\bar{K}>0$ such that,
\begin{align*}
\EE|\frac1N\|\x_j\|^2-1|^6 &\leq 
KN^{-3}\log^{12}N \\
\EE\frac1{N^6} |\x_j^\herm \R^\oh(\B_{(j)}-z\I_N)^{-1}\E\D_{(j)}^{-1}\R^\oh\x_j
 -\tr \R(\B_{(j)}-z\I_N)^{-1}\E\D_{(j)}^{-1}|^6 &\leq 
 KN^{-3}v^{-12}\log^{24}N \\
 \EE \frac1{N^6}|\x_j^\herm\R^\oh(\B_{(j)}-z\I_N)^{-1}\R^\oh \x_j 
 -\tr\R^\oh(\B_{(j)}-z\I_N)^{-1}\R^\oh|^6 &\leq KN^{-3}v^{-6}\log^{18}N.
 \end{align*}
All three moments when multiplied by $n$ times any power of $\log N$, are summable. Applying standard arguments using the Borel-Cantelli lemma and Boole's inequality (on $4n$ events), we conclude that, for any $k>0$
$\log^kN\max_{j\leq n}\tau_jd_j\asto 0$ as $N\to\infty$. Hence Equations \eqref{eq:9} and \eqref{eq:9b}. 

\subsection{Existence and uniqueness of $m_N^\circ(z)$}
We show now that for any $N$, $n$, $\S$, $\R$, $N\times N$ nonnegative definite and $\T=\diag(\tau_1,\ldots,\tau_N)$, $\tau_k\geq 0$ for all $1\leq k\leq N$, there exists a unique $e$ with positive imaginary part for which 
\begin{equation}
  \label{eq:10}
  e=\frac1N\tr\left(\S+\left[\int\frac{\tau}{1+c_N\tau e}dF^{\T}(\tau)\right]\R-z\I_N\right)^{-1}\R.
\end{equation}
For existence we consider the subsequences $\{N_j\}$, $\{n_j\}$ with $N_j=jN$, $n_j=jn$, so that $c_{N_j}$ remains $c_N$, form the block diagonal matrices 
\begin{equation}
  \R_{N_j}=\diag(\R,\R,\ldots,\R),\quad \S_{N_j}=\diag(\S,\S,\ldots,\S)
\end{equation}
both $jN\times jN$ and 
\begin{equation}
\T_{N_j}=\diag(\T,\T,\ldots,\T)
\end{equation}
of size $jn\times jn$. We see that $F^{\T_{N_j}}=F^{\T}$ and the right side of \eqref{eq:10} remains unchanged for all $N_j$. Consider a realisation where $w_{N_j}^e\to 0$ as $j\to\infty$. We have $|e_{N_j}(z)|=|(jN)^{-1}\tr \R(\B_{jN}-z\I_N)^{-1}|\leq v^{-1}\log N$, remaining bounded as $j\to\infty$. Consider then a subsequence for which $e_{N_j}$ converges to, say, $e$. From \eqref{eq:6}, we see that 
\begin{equation*}
\left|\frac{\tau}{1+c_N\tau e_{N_j}}\right|\leq \tau|z|v^{-1}
\end{equation*}
so that from the dominated convergence theorem we have
\begin{equation*}
\int\frac{\tau}{1+c_N\tau e_{N_j}(z)}dF^{\T}(\tau)\to \int\frac{\tau}{1+c_N \tau e}dF^{\T}(\tau)
\end{equation*}
along this subsequence. Therefore $e$ solves \eqref{eq:10}.

We now show uniqueness. Let $e$ be a solution to \eqref{eq:10} and let $e_2=\Im[e]$.
Recalling the definition of $\D$ we write
\begin{equation*}
  e=\frac1N \tr \left(\D^{-1}\R\D^{-\herm}\left(\S + \left[ \int \frac{\tau}{1+c_N\tau e^\ast}dF^\T(\tau) \right]\R - z^\ast\I \right) \right).
\end{equation*}
We see that since both $\R$ and $\S$ are Hermitian nonnegative definite, $\tr \left(\D^{-1}\R\D^{-\herm}\S\right)$ is real and nonnegative. Therefore we can write
\begin{equation}
  \label{eq:e2K1}
  e_2=\frac1N\tr\left(\D^{-1}\R(\D^{\herm})^{-1}\left(\left[\int\frac{c_N\tau^2e_2}{|1+c_N\tau e|^2}dF^{\T}(\tau)\right]\R+v\I_N\right)\right)=e_2\alpha+v\beta,
\end{equation}
where we denoted 
\begin{align*}
  \alpha &=\frac1N\tr\left(\D^{-1}\R(\D^{\herm})^{-1}\left[\int\frac{c_N\tau^2}{|1+c_N\tau e|^2}dF^{\T}(\tau)\right]\R\right) \\
  \beta &= \frac1N\tr\left(\D^{-1}\R(\D^{\herm})^{-1}\right).
\end{align*}
Let $\underline e$ be another solution to \eqref{eq:10}, with $\underline e_2=\Im[\underline e]$, and analogously we can write $\underline e_2=\underline e_2\underline\alpha+v\underline\beta$. Let $\underline \D$ denote $\D$ with $e$ replaced by $\underline e$.
Then we have $e-\underline e=\gamma(e-\underline e)$ where 
\begin{equation*}
\gamma=\int\frac{c_N\tau^2}{(1+c_N\tau e)(1+c_N\tau\underline e)}dF^{\T}(\tau)
\frac{\tr \D^{-1}\R\underline \D^{-1}\R}{N}.
\end{equation*}
If $\R$ is the zero matrix, then $\gamma=0$, and $e=\underline e$ would follow. For $\R\neq 0$ we use Cauchy-Schwarz to find 
\begin{align*}
|\gamma| &\leq \left(\int\frac{c_N\tau^2}{|1+c_N\tau e|^2}dF^{\T}(\tau)
\frac{\tr \D^{-1}\R(\D^{\herm})^{-1}\R}{N}\right)^\oh
\left(\int\frac{c_N\tau^2}{|1+c_N\tau\underline e|^2}dF^{\T}(\tau)
\frac{\tr\underline \D^{-1}\R({\underline \D}^{\herm})^{-1}\R}{N}\right)^\oh \\
&=\alpha^\oh\underline\alpha^\oh \\
&=\left(\frac{e_2\alpha}{e_2\alpha+v\beta}\right)^\oh
\left(\frac{\underline e_2\underline\alpha}{\underline e_2\underline\alpha+v\underline\beta}\right)^\oh.
\end{align*}
Necessarily $\beta$ and $\underline\beta$ are positive since $\R\neq0$. Therefore $|\gamma|<1$ so we must have $e=\underline e$. For $z<0$ and $e>0$, the same calculus can be performed, with $\gamma$ remaining the same. The step \eqref{eq:e2K1} is changed by evaluating $e$, instead of $e_2$, using the same technique. We obtain the same $\alpha$ while $\beta$ is replaced by another positive scalar. We therefore still have that $\gamma<1$.

\subsection{Termination of the proof}
Let $e_N^\circ $ denote the solution to \eqref{eq:10}. We show now for any $\ell>0$, almost
surely
\begin{equation}
  \label{eq:11}
  \lim_{N\to\infty}\log^\ell N(e_N-e_N^\circ )=0.
\end{equation}
Let $e_2^\circ =\Im [e_N^\circ ]$, and  $\alpha^\circ =\alpha_N^\circ $, $\beta^\circ =\beta^\circ _N$ be the values as above for which $e_2^\circ =e_2^\circ \alpha^\circ +v\beta^\circ $.
We have, using \eqref{eq:3} and \eqref{eq:6},
\begin{align*}
e_2^\circ \alpha_N^\circ /\beta_N^\circ  &\leq e_2^\circ c_N\log N\int\frac{\tau^2}{|1+c_N\tau e_N^\circ |^2}dF^{\T}(\tau) \\
&=-\log N\Im\left[\int\frac{\tau}{1+c_N\tau e_N^\circ }dF^{\T}(\tau)\right] \\ 
&\leq\log^2 N|z|v^{-1}.
\end{align*}
Therefore
\begin{align}
  \label{eq:12}
\alpha^\circ  &=\left(\frac{e_2^\circ \alpha^\circ }{e_2^\circ \alpha^\circ +v\beta^\circ }\right) \nonumber \\
&=\left(\frac{e_2^\circ \alpha^\circ /\beta^\circ }{v+e_2^\circ \alpha^\circ /\beta^\circ }\right) \nonumber \\
&\leq\left(\frac{\log^2 N|z|}{v^2+\log^2 N|z|}\right)
\end{align}
Let $\D^\circ $, $\D$ denote $\D$ as above with $e$ replaced by, respectively $e_N^\circ $ and $e_N$.  We have 
\begin{equation*}
  e_N=\frac1N\tr \D^{-1}\R-w_N^e.
\end{equation*}
With $e_2=\Im [e_N]$ we write as above
\begin{align*}
  e_2&=\frac1N\tr\left(\D^{-1}\R\D^{-\herm}\left(\left[\int\frac{c_N\tau^2e_2}{|1+c_N\tau e_N|^2}dF^{\T}(\tau)\right]\R+v\I_N\right)\right) -\Im [w_N^e] \\
&=e_2\alpha+v\beta-\Im w_N^e
\end{align*}
We have as above $e_N-e_N^\circ =\gamma(e_N-e_N^\circ )+w_N^e$ where now
\begin{equation*}
|\gamma|\leq {\alpha^\circ }^\oh\alpha^\oh.
\end{equation*}
Fix an $\ell>0$ and consider a realisation for which $\log^{\ell'}N\,w_N^e\to0$, where
$\ell'=\max(\ell+1,4)$ and $n$ large enough so that
\begin{equation}
  \label{eq:13}
|w_N^e|\leq\frac{v^3}{4c_N|z|^2\log^3N}.
\end{equation}
Suppose $\beta\leq\frac{v^2}{4c_N|z|^2\log^3N}$. Then by Equations \eqref{eq:3} and \eqref{eq:6} we get
\begin{equation*}
\alpha\leq c_Nv^{-2}|z|^2\log^3N\beta\leq 1/4,
\end{equation*}
which implies $|\gamma|\leq 1/2$. Otherwise we get from \eqref{eq:12} and \eqref{eq:13}
\begin{align*}
|\gamma| &\leq{\alpha^\circ }^\oh\left(\frac{e_2\alpha}{e_2\alpha+v\beta-\Im[w_N^e]}\right)^\oh\\
& \leq\left(\frac{\log N|z|}{v^2+\log N|z|}\right)^\oh.
\end{align*}
Therefore for all $N$ large
\begin{align*}
\log^\ell N|e_N-e_N^\circ | &\leq\frac{(\log^\ell N)w_N^e}{1-\left(\frac{\log^2 N|z|}{v^2+\log^2 N|z|}\right)^\oh} \\
&\leq 2v^{-2}(v^2+\log^2 N|z|)(\log^\ell N)w_N^e\\
&\to 0,
\end{align*}
as $n\to\infty$. Therefore \eqref{eq:11} follows.

Let $m_N^\circ =N^{-1}\tr \D^\circ $. We finally show
\begin{equation}
  \label{eq:14}
m_N-m_N^\circ \asto 0,
\end{equation}
as $n\to\infty$. Since $m_N=N^{-1}\tr \D^{-1}-w_N^m$, we have 
\begin{equation*}
m_N-m_N^\circ =\gamma(e_N-e_N^\circ ) -w_N^m,
\end{equation*}
where now 
\begin{equation*}
\gamma=\int\frac{c_N\tau^2}{(1+c_N\tau e_N)(1+c_N\tau e_N^\circ )}dF^{\T}(\tau)
\frac{\tr \D^{-1}\R{\D^\circ }^{-1}}N.
\end{equation*}
From \eqref{eq:3} and \eqref{eq:6} we get $|\gamma|\leq c_N|z|^2v^{-4}\log^3N$.
Therefore, from \eqref{eq:9} and \eqref{eq:11}, we get \eqref{eq:14}.

Returning to the original assumptions on $X_{11}$, $\T$, and $\R$, for each of a countably infinite collection of $z$ with positive imaginary part, possessing a cluster point with positive imaginary part, we have \eqref{eq:14}. Therefore, by Vitali's convergence theorem, page 168 of \cite{TIT39}, for any $\varepsilon>0$ we have with probability one $m_N(z)-m_N^\circ (z)\to 0$ uniformly in any region of $\CC$ bounded by a contour interior to 
\begin{equation*}
\CC\setminus \left(\{z:|z|\leq\epsilon\}\cup\{z=x+iv:x>0,|v|\leq\varepsilon\}\right).
\end{equation*}

If $\S=f(\R)$, meaning the eigenvalues of $\R$ are changed via $f$ in the spectral decomposition of $\R$, then we have
\begin{align*}
m_N^\circ (z) &=\int\frac1{f(r)+r\int\frac{\tau}{1+c_N\tau e^\circ _N(z)}dF^{\T}(\tau)-z}dF^{\R}(r) \\
e_N^\circ (z) &=\int\frac r{f(r)+r\int\frac{\tau}{1+c_N\tau e_N^\circ (z)}dF^{\T}(\tau)-z}dF^{\R}(r).
\end{align*}

\subsection{Extension to $K\geq 1$}
  \label{sec:proof_K}
Suppose now 
\begin{equation*}
  \B_N=\S+\sum_{k=1}^K\R_k^\oh\X_k\T_k\X_k^\herm\R_k^\oh
\end{equation*}
where $K$ remains fixed, $\X_k$ is $N\times n_k$ satisfying \ref{item:X}, the $\X_k$'s are independent, $\R_k$ satisfies \ref{item:R}) and \ref{item:tight}), $\T_k$ is $n_k\times n_k$ satisfying \ref{item:T}) and \ref{item:tight}), $c_k = N/n_k$ satisfies \ref{item:c}), and $\S$ satisfies \ref{item:S}). After truncation and centralisation we may assume the same condition on the entries of the $\X_k$'s, and the spectral norms of the $\R_k$'s and the $\T_k$'s. Write $\y_{k,j}=(1/\sqrt{n_k})\R_k^\oh \x_{k,j}$, with $\x_{k,j}$ denoting the $j$-th column of $\X_k$, and let $\tau_{k,j}$ denote the $j$-th diagonal element of $\T_k$. Then we can write
\begin{equation*}
\B_N=\S+\sum_{k=1}^K\sum_{j=1}^{n_k}\tau_{k,j}\y_{k,j}\y_{k,j}^\herm.
\end{equation*}
Define
\begin{equation*}
  e_{N,k}=e_{N,k}(z)=(1/N)\tr \R_k(\B_N-z\I_N)^{-1}
\end{equation*}
and
\begin{align*}
  p_k &= -\frac1{n_kz}\sum_{j=1}^{n_k}\frac{\tau_{k,j}}{1+c_k\tau_{k,j}e_{N,k}} \\
  &= \int\frac{-\tau_k}{1+c_k\tau_k e_{N,k}}dF^{\T_k}(\tau_k).
\end{align*}
We see $e_{N,k}$ and $p_k$ have the same properties as $e_N$ and $p_N$. Let $\B_{k,(j)}=\B_N-\tau_{k,j}\y_{k,j}\y_{k,j}^\herm$. Define $\D=-z\I_N+\S-\sum_{k=1}^K zp_k(z)\R_k$. We write
\begin{align*}
 \B_N-z\I_N -\D =\sum_{k=1}^K\left(\sum_{j=1}^{n_k}\tau_{k,j}\y_{k,j}\y_{k,j}^\herm+zp_k(z)\R_k\right).
\end{align*}
Taking inverses and using Lemma \ref{le:4}, we have
\begin{align*}
\D^{-1}-(\B_N-z\I_N)^{-1}&=\sum_{k=1}^K\left(\sum_{j=1}^{n_k}\tau_{k,j}\D^{-1}
\y_{k,j}\y_{k,j}^\herm(\B_N-z\I_N)^{-1}+zp_k\D^{-1}\R_k(\B_N-z\I_N)^{-1}\right) \\
&=\sum_{k=1}^K\left(\sum_{j=1}^{n_k}\tau_{k,j}\frac{\D^{-1}\y_{k,j}\y_{k,j}^\herm
(\B_{k,(j)}-z\I_N)^{-1}}{1+\tau_{k,j}\y_{k,j}^\herm(\B_{k,(j)}-z\I_N)^{-1}\y_{k,j}}+
zp_k\D^{-1}\R_k(\B_N-z\I_N)^{-1}\right).
\end{align*}
Taking traces and dividing by $N$, we have
\begin{equation*}
  (1/N)\tr \D^{-1}-m_N(z)=\sum_{k=1}^K\frac1{n_k}\sum_{j=1}^{n_k}\tau_{k,j}
d_{k,j}\equiv w_N^m,
\end{equation*}
where
\begin{equation*}
d_{k,j}=\frac{(1/N)\x_{k,j}^\herm \R_k^\oh(\B_{k,(j)}-z\I_N)^{-1}\D^{-1}
\R_k^\oh \x_{k,j}}{1+\tau_{k,j}\y_{k,j}^\herm(\B_{k,(j)}-z\I_N)^{-1}\y_{k,j}}
-\frac{(1/N)\tr \R_k(\B_N-z\I_N)^{-1}\D^{-1}}{1+c_k\tau_{k,j}e_{N,k}}.
\end{equation*}
For a fixed $\underline k\in\{1,\ldots,K\}$, we multiply the above matrix identity by $\R_{\underline k}$, take traces and divide by $N$. Thus we get
\begin{equation*}
  (1/N)\tr \D^{-1}\R_{\underline k}- e_{\underline k}(z)=\sum_{k=1}^K\frac1{n_k}\sum_{j=1}^{n_k}\tau_{k,j}d_{k\underline{k}j}^e\equiv w_{\underline k}^e,
\end{equation*}
where 
\begin{equation*}
  d^e_{k\underline{k}j}=\frac{(1/N)\x_{k,j}^\herm\R_k^\oh(\B_{k,(j)}-z\I_N)^{-1}\R_{\underline k}\D^{-1}\R_k^\oh\x_{k,j}}{1+\tau_{k,j}\y_{k,j}^\herm(\B_{k,(j)}-z\I_N)^{-1}\y_{k,j}}
-\frac{(1/N)\tr \R_k(\B_N-z\I_N)^{-1}\R_{\underline k}\D^{-1}}{1+c_k\tau_{k,j}e_{N,k}}.
\end{equation*}

In exactly the same way as in the case with $K=1$ we find that for any nonnegative $\ell$, $\log^\ell Nw_N^m$ and the $\log^\ell w_{\ui}^e$'s converge almost surely to zero. By considering block diagonal matrices as before with $N$, $n_i$'s, $\S$, $\R_i$'s and $\T_i$'s all fixed we find that there exist $e_1^\circ ,\ldots,e_K^\circ $ with positive imaginary parts for which for each $i$
\begin{equation}
  \label{eq:15}
e_i^\circ =\frac1N\tr\R_i \left(\S+\sum_{k=1}^K\left[\int\frac{\tau}{1+c_k\tau e_k^\circ }dF^{\T_k}(\tau)\right]\R_k-z\I_N\right)^{-1}.
\end{equation}

Let us verify uniqueness. Let $\e^\circ =(e_1^\circ ,\ldots,e_K^\circ )^\trans$, and let $\D^\circ $ denote the matrix in \eqref{eq:15} whose inverse is taken (essentially $\D$ after the $e_{N,i}$'s are replaced by the $e_i^\circ $'s). Let for each $j$, $e_{j,2}^\circ =\Im e_j^\circ $, and $\e_2^\circ =(e_{1,2}^\circ ,\ldots,e_{K,2}^\circ )^\trans$. 
Then, noticing that for each $i$, $\tr \S{\D^\circ }^{-\herm}\R_i{\D^\circ }^{-1}$ is real and nonnegative (positive whenever $\S\neq 0$) and $\tr{\D^\circ }^{-\herm}\R_i{\D^\circ }^{-1}$ and $\tr \R_j{\D^\circ }^{-\herm}\R_i{\D^\circ }^{-1}$ are real and positive for all $i$, $j$, we have 
\begin{align}
	\label{eq:ei2}
  e_{i,2}^\circ  &= \Im\left[\frac1N\tr \left(\S+\sum_{j=1}^K\left[\int\frac{\tau}{1+c_j\tau\overline e_j^\circ }dF^{\T_j}(\tau)\right]\R_j-z^\ast \I\right){\D^\circ }^{-\herm}\R_i{\D^\circ }^{-1}\right] \nonumber \\
  &=\sum_{j=1}^Ke_{j,2}^\circ  \frac1N\tr \R_j{\D^\circ }^{-\herm}\R_i{\D^\circ }^{-1}c_j\int\frac{\tau^2}{|1+c_j \tau e_j^\circ |^2}dF^{\T_j}(\tau)+\frac{v}N\tr{\D^\circ }^{-\herm}\R_i{\D^\circ }^{-1}.
\end{align}

Let $\CCC^\circ =(c_{ij}^\circ )$, ${\bf b}^\circ =(b_1^\circ ,\ldots,b_N^\circ )^\trans$, where
\begin{equation*}
  c_{ij}^\circ =\frac1N\tr \R_j{\D^\circ }^{-\herm}\R_i{\D^\circ }^{-1}c_j\int\frac{\tau^2}{|1+c_j\tau e_j^\circ |^2}dF^{\T_j}(\tau)
\end{equation*}
and
\begin{equation*}
  b_i^\circ =\frac1N\tr{\D^\circ }^{-\herm}\R_i{\D^\circ }^{-1}.
\end{equation*}
Therefore we have that $\e_2^\circ $ satisfies
\begin{equation}
  \label{eq:16}
  \e_2^\circ =\CCC^\circ \e_2^\circ +v{\bf b}^\circ .
\end{equation}
We see that each $e_{j,2}^\circ $, $c_{ij}^\circ $, and $b_j^\circ $ are positive. 
Therefore, from Lemma \ref{le:7} we have $\rho(\CCC^\circ )<1$.

Let $\underline\e^\circ =(\underline e_1^\circ ,\ldots,\underline e_K^\circ )^\trans$ be another solution to \eqref{eq:15}, with $\underline\e^\circ _2$, $\underline \D^\circ $, $\underline \CCC^\circ =(\underline c_{ij}^\circ )$, $\underline{\bf b}^\circ $ defined analogously, so that \eqref{eq:16} holds and $\rho(\underline \CCC^\circ )<1$. We have for each $i$,
\begin{equation*}
e_i^\circ -\underline e_i^\circ =\frac1N\tr \R_i{\D^\circ }^{-1}\sum_{j=1}^K(e_j^\circ -\underline e_j^\circ )c_j\int\frac{\tau^2}{(1+c_j\tau e_j^\circ )(1+c_j\tau\underline e_j^\circ )}dF^{\T_j}(\tau)\R_j{\underline \D^\circ }^{-1}.
\end{equation*}
Thus with $\A=(a_{ij})$ where
\begin{equation}
	\label{eq:aij}
a_{ij}=\frac1N\tr \R_i{\D^\circ }^{-1}\R_j{\underline \D^\circ }^{-1}c_j\int\frac{\tau^2}{(1+c_j\tau e_j^\circ )(1+c_j\tau\underline e_j^\circ )}dF^{\T_j}(\tau),
\end{equation}
we have 
\begin{equation}
	\label{eq:e0me}
  \e^\circ -\underline\e^\circ =\A(\e^\circ -\underline\e^\circ),
\end{equation}
which means, if $\e^\circ \neq \underline\e^\circ $, then $\A$ has an eigenvalue equal to 1.

Applying Cauchy-Schwarz we have
\begin{align*}
  |a_{ij}|&\leq \left(\frac1N \R_i{\D^\circ }^{-1}\R_j{\D^\circ }^{-\herm}
\int\frac{\tau^2}{|1+c_j\tau e_j^\circ |^2}dF^{\T_j}(\tau)\!\right)^\oh
\left(\frac1N \R_i{\underline \D^\circ }^{-1}\R_j{{\underline \D}^\circ }^{-\herm}
\int\frac{\tau^2}{|1+c_j\tau \underline e_j^\circ |^2}dF^{\T_j}(\tau)\right)^\oh \\
&={c_{ij}^\circ }^{1/2}{\underline c_{ij}^\circ }^{1/2}.
\end{align*}

Therefore from Lemmas \ref{le:8} and \ref{le:9} we get
\begin{equation*}
\rho(\A)\leq\rho({c_{ij}^\circ }^\oh{\underline c_{ij}^\circ }^\oh)\leq \rho(\CCC^\circ )^\oh\rho(\underline \CCC^\circ )^\oh<1;
\end{equation*}
a contradiction to the statement $\A$ has an eigenvalue equal to 1. Consequently we have $\e=\underline\e$.

The same reasoning can be applied to $z<0$, with $e_i^\circ >0$. In this case matrix $\A$ remains the same. The step \eqref{eq:ei2} is now replaced by taking $e_i^\circ $, instead of its imaginary part, using the same line of reasoning. This leads to the same matrix $\CCC^\circ$ with \eqref{eq:16} remaining true with $\b^\circ $ replaced by another positive vector. The conclusion $\rho(\A)<1$ therefore remains.

Let $\e_N=(e_{N,1},\ldots,e_{N,K})^\trans$ and $\e_N^\circ =(e_{N,1}^\circ ,\ldots,
e_{N,K}^\circ )$ denote the vector solution to \eqref{eq:15} for each $N$. We will show for any $\ell>0$, almost surely
\begin{equation}
  \label{eq:17}
  \lim_{N\to\infty}\log^{\ell}N(\e_N-\e_N^\circ )\to {\bf 0}.
\end{equation}
We have
\begin{equation*}
  \e_N^\circ =(\frac1N\tr \R_1{\D^\circ }^{-1},\ldots,\frac1N\tr \R_K{\D^\circ }^{-1})^\trans.
\end{equation*}
Let $\w^e=\w^e_N=-(w_1^e,\ldots,w_K^e)^\trans$. Then we can write
\begin{equation*}
  \e_N=(\frac1N\tr \R_1\D^{-1},\ldots,\frac1N\tr \R_K\D^{-1})^\trans +\w^e.
\end{equation*}
Therefore
\begin{equation*}
  \e_N-\e_N^\circ =\A(N)(\e_N-\e_N^\circ )+\w^e,
\end{equation*}
where $\A(N)=(a_{ij}(N))$ with 
\begin{equation*}
a_{ij}(N)=\frac1N\tr \R_i\D^{-1}\R_j{\D^\circ }^{-1}c_j\int\frac{\tau^2}{(1+c_j\tau e_{N,j})(1+c_j\tau e_{N,j}^\circ )}dF^{\T_j}(\tau).
\end{equation*}
We let $\e_{N,2}^\circ $, $b_{ij}^\circ (N)$, $\CCC^\circ (N)$, $b_{N,i}^\circ $, and ${\bf b}_N^\circ $, denote the quantities from above, reflecting now their dependence on $N$. 
Let $\CCC(N)=(c_{ij}(N))$ be $K\times K$ with 
\begin{equation*} 
  c_{ij}(N)=\frac1N\tr \R_j\D^{-\herm}\R_i \D^{-1}c_j\int\frac{\tau^2}{|1+c_j \tau e_{N,j}|^2}dF^{\T_j}(\tau).
\end{equation*}

Let $\e_{N,2}=\Im[\e_N]$ and $\w^e_2=\Im[\w^e]$. Define ${\bf b}_N=(b_{N,1},\ldots,b_{N,K})^\trans$ with
\begin{equation*}
  b_{N,i}=\frac1N\tr \D^{-\herm}\R_i\D^{-1}.
\end{equation*}
Then, as above we find that
\begin{equation}
  \label{eq:18}
  \e_{N,2}=\CCC(N)\e_{N,2}+v{\bf b}_N+\w^e_2.
\end{equation}

Using \eqref{eq:3} and \eqref{eq:6} we see there exists a constant $K_1>0$ for which
\begin{align*}
c_{ij}^\circ (N)\leq K_1\log^3N b_{N,i}^\circ 
\end{align*}
and
\begin{align*}
c_{ij}(N) &\leq K_1\log^3N b_{N,i} \\
c_{ij}(N) &\leq K_1\log^4N
\end{align*}
for each $i,j$. Therefore, from \eqref{eq:16} we see there exists $\hat{K}>0$ for which
\begin{equation}
  \label{eq:19}
  e_{N,i}^\circ \leq \hat{K}\log^4N vb_{N,i}^\circ.
\end{equation}
Let $\x$ be such that $\x^\trans$ is a left eigenvector of $\CCC^\circ (N)$ corresponding to eigenvalue $\rho(\CCC^\circ (N))$, guaranteed by Lemma \ref{le:10}. Then from \eqref{eq:18} we have
\begin{equation}
  \label{eq:20}
  \x^\trans\e_{N,2}^\circ =\rho(\CCC^\circ (N))\x^\trans\e_{N,2}^\circ +v\x^\trans {\bf b}_N^\circ .
\end{equation}
Using \eqref{eq:20} we have
\begin{equation}
  \label{eq:21}
  1-\rho(\CCC^\circ (N))=\frac{v\x^\trans{\bf b}_N^\circ }{\x^\trans\e^\circ _{N,2}}\ge (\hat{K}\log^4N)^{-1}.
\end{equation}

Fix an $\ell>0$ and consider a realisation for which $\log^{\ell+3+p}N\w^e_N\to 0$, as $N\to\infty$, where $p\geq 12K-7$. We will show for all $N$ large
\begin{equation}
  \label{eq:22}
  \rho(\CCC(N))\leq 1+(\hat{K}\log^4N)^{-1}.
\end{equation}

For each $N$ we rearrange the entries of $\e_{N,2}$, 
$v{\bf b}_m+\w^e_2$, and  $\CCC(n)$ depending on whether
the $i^{th}$ entry of $v{\bf b}_m+\w^e_2$ is greater than, or less
than or equal to zero. We can therefore assume
\begin{equation*}
  \CCC=\begin{pmatrix} \CCC_{11}(N) & \CCC_{12}(N)\\ \CCC_{21}(N) & \CCC_{22}(N)\end{pmatrix},
\end{equation*}
where $\CCC_{11}(N)$ is $k_1\times k_1$, $\CCC_{22}(N)$ is $k_2\times k_2$,
$\CCC_{12}(N)$ is $k_1\times k_2$, and $\CCC_{21}(N)$ is $k_2\times k_1$.
From Lemma \ref{le:7} we have $\rho(\CCC_{11}(N))<1$. If $vb_{N,i}+\w^e_{2,i}\leq0$,
then necessarily $vb_{N,i}\leq|\w^e_N|\leq K_1(\log n)^{-(3+p)}$, and
so from \eqref{eq:19} we have the entries of $\CCC_{21}(N)$ and $\CCC_{22}(N)$ bounded by
$K_1(\log N)^{-p}$.  We may assume for all $N$ large $0<k_1<K$, since
otherwise we would have $\rho(\CCC(N))<1$.   

We seek an expression for $\det(\CCC(N)-\lambda \I_N)$ in which Lemma \ref{le:12} can be
used. We consider $N$ large enough so that, for $|\lambda|\ge 1/2$, we
have $(\CCC_{22}(N)-\lambda \I_N)^{-1}$ existing with entries uniformly bounded.
We have
\begin{align*}
  \det(\CCC(N)-\lambda \I) &= \det\left[\begin{pmatrix} \I&-
\CCC_{12}(N)(\CCC_{22}(N)-\lambda \I)^{-1}\\
0&\I \end{pmatrix}\begin{pmatrix} \CCC_{11}(N)-\lambda \I &\CCC_{12}(N)\\ \CCC_{21}(N)&
  \CCC_{22}(N)-\lambda \I\end{pmatrix}\right] \\
&=\det\begin{pmatrix} \CCC_{11}(N)-\lambda \I-\CCC_{12}(N)(\CCC_{22}(N)-\lambda \I)^{-1}
\CCC_{21}(N)&0\\ \CCC_{21}(N) & \CCC_{22}(N)-\lambda \I\end{pmatrix} \\
&=\det(\CCC_{11}(N)-\lambda \I-\CCC_{12}(N)(\CCC_{22}(N)-\lambda \I)^{-1}\CCC_{21}(N))
\det(\CCC_{22}(N)-\lambda \I).
\end{align*}

We see then that for $\lambda=\rho(\CCC(N))$ real and greater than $1$, 
\begin{equation}
  \label{eq:23}
\det(\CCC_{11}(N)-\lambda \I-\CCC_{12}(N)(\CCC_{22}(N)-\lambda \I)^{-1}\CCC_{21}(N))
\end{equation}
must be zero. 

Notice that from \eqref{eq:19}, the entries of $\CCC_{12}(N)(\CCC_{22}(N)-\lambda \I)^{-1}\CCC_{21}(N)$ can be made smaller than any negative power of $\log N$ for $p$ sufficiently large.
Notice also that the diagonal elements of $\CCC_{11}(N)$ are all less than $1$. 
From this, Lemma \ref{le:11} and \eqref{eq:19}, we see that $\rho(\CCC(N))\leq K_1\log^4N$.
The determinant in \eqref{eq:23} can be written as
\begin{equation*}
\det(\CCC_{11}(N)-\lambda \I)+g(\lambda),
\end{equation*}
where $g(\lambda)$ is a sum of products, each containing at least one
entry from $\CCC_{12}(N)(\CCC_{22}(N)-\lambda \I)^{-1}\CCC_{21}(N)$. Again, from \eqref{eq:19} we see that for all $|\lambda|\ge1/2$, $g(\lambda)$ can be made smaller than any negative
power of $\log N$ by making $p$ sufficiently large. Choose $p$ so that $|g(\lambda)|<(\widehat K\log N)^{-4k_1}$ for these $\lambda$. It is clear that any
$p>8k_1+4$ will suffice. Let $\lambda_1,\ldots,\lambda_{k_1}$ denote the eigenvalues of $\CCC_{11}$. Since $\rho(\CCC_{11})<1$, we see that for $|\lambda|\ge(\widehat K\log N)^{-4}$, we have
\begin{align*}
|\det(\CCC_{11}(N)-\lambda \I)|&=|\prod_{i=1}^{k_1}(\lambda_i-\lambda)| \\
& > (\widehat K\log N)^{-4k_1}.
\end{align*}
Thus with $f(\lambda)=\det(\CCC_{11}(N)-\lambda \I)$, a polynomial, and
$g(\lambda)$ being a rational function, we have the conditions of
Lemma \ref{le:12} being met on any rectangle $C$, with vertical lines going through
$((\widehat K\log N)^{-4},0)$ and $(K_1(\log N)^4,0)$. Therefore, since
$f(\lambda)$ has no zeros inside $C$, neither does $\det(\CCC(N)-\lambda \I)$.  
Thus we get \eqref{eq:22}. As before we see that
\begin{equation*}
|a_{ij}(N)|\leq c_{ij}^{1/2}(N){c_{ij}^\circ }^{1/2}(N).
\end{equation*}
Therefore, from \eqref{eq:21}, \eqref{eq:22}, and Lemmas \ref{le:8} and \ref{le:9}, we have for all $N$ large
\begin{equation}
  \label{eq:24}
\rho(\A(N))\leq\left(\frac {\widehat K^2\log^8N-1}{\widehat K^2\log^8N}\right)^\oh.
\end{equation}
For these $N$ we have then $\I-\A(N)$ invertible, and so
\begin{equation*}
\e_N-\e_N^{\,0}=(\I-\A(N))^{-1}\w^e.
\end{equation*}
By \eqref{eq:3} and \eqref{eq:6} we have the entries of $\A(N)$ bounded by $K_1\log^4N$.  
Notice also, from \eqref{eq:24}
\begin{equation*}
|\det(\I-\A(N))|\ge(1-\rho(\A(N))^K\ge\left(\widehat K^2\log^8N
\left(1+\frac {\widehat K^2\log^8N-1}
{\widehat K^2\log^8N}\right)^\oh\right)^{-K}\ge
(2\widehat K^2\log^8N)^{-K}.
\end{equation*}

When considering the inverse of a square matrix in terms of its adjoint 
divided by its determinant, we see that the entries of $(\I-\A(N))^{-1}$ are
bounded by 
\begin{equation*}
\frac{(K-1)!K_1(\log N)^{4(K-1)}}{|\det(\I-\A(N))|}\leq K_3(\log N)^{12K-4}.
\end{equation*}
Therefore, since $p\ge 12K-7$ ($>8k_1+4$), \eqref{eq:17} follows on this 
realisation, an event which occurs with probability one. 

Letting $m_N^\circ =\frac1N\tr{\D^\circ }^{-1}$, we have
\begin{equation*}
m_N-m_N^\circ =\vec\gamma^\trans(\e_N-\e_N^\circ ),
\end{equation*}
where $\vec\gamma=(\gamma_1,\ldots,\gamma_K)^\trans$ with
\begin{equation*}
\gamma_j=\int\frac{c_N\tau^2}{(1+c_N\tau e_{N,j})(1+c_n\tau e_{N,j}^\circ )}dF^{\T_N}(\tau)\frac{\tr \D^{-1}\R_j{\D^\circ }^{-1}}N.
\end{equation*}
From \eqref{eq:3} and \eqref{eq:6} we get each $|\gamma_j|\leq c_N|z|^2v^{-4}\log^3N$.
Therefore from \eqref{eq:17} and the fact that $w_N^m\to 0$, we have 
\begin{equation*}
  m_N-m_N^\circ \to 0,
\end{equation*}
almost surely, as $N\to\infty$.

This completes the proof.

\section{Proof of Theorem \ref{th:2}}
\label{ap:th2}
We first prove that $\mathcal V_N^\circ(x)$ as defined in Equation \eqref{eq:th2V2} verifies
\begin{equation} 
  \label{eq:proof21}
  \mathcal V_N^\circ(x) = \int_x^{\infty} \left( \frac1w - m_N^\circ(-w) \right) dw
\end{equation}
and then we prove that, under the conditions of Theorem \ref{th:2}, $\mathcal V^\circ (x)$ defined as such verifies
\begin{equation}
  \label{eq:proof22}
  \mathcal V_N^\circ(x) - \mathcal V_N(x) \asto 0.
\end{equation}

\subsection{Proof of \eqref{eq:proof21}}
First, write $e_i(z)$ under the symmetric form
\begin{align*}  
  e_i(z) &= \frac1N \tr \R_i\left(-z \left[\I_N + \sum_{k=1}^K\delta_k\R_k\right] \right)^{-1} \\
  \delta_i(z) &= \frac1{n_i} \tr \T_i \left(-z \left[\I_{n_i} + c_ie_i(z)\T_i \right] \right)^{-1}
\end{align*}
and then for $m_N^\circ(z)$,
\begin{equation*}
  m_N^\circ(z) = \frac1N \tr \left(-z \left[\I_N + \sum_{k=1}^K\delta_k\R_k\right] \right)^{-1}.
\end{equation*}

Now, notice that 
\begin{align*}
  \frac1z - m_N^\circ(-z) &= \frac1N\left( \left(z\I\right)^{-1}-\left(z\left[\I_N + \sum_{k=1}^K \delta_k\R_k\right]\right)^{-1}\right) \\
  &= \sum_{k=1}^K \delta_k(-z)\cdot e_k(-z).
\end{align*}
Since the Shannon transform $\mathcal V(x)$ satisfies $\mathcal V(x)=\int_x^{+\infty}[w^{-1}- m_N(-w)] dw$, we need to find an integral form for $\sum_{k=1}^K \delta_k(-z)\cdot e_k(-z)$.
Notice now that
\begin{align*}
  \frac{d}{dz}\frac1N\log\det\left(\I_N + \sum_{k=1}^K \delta_k(-z) \R_k \right) &= - z \sum_{k=1}^K e_k(-z)\cdot \delta_k'(-z) \nonumber \\
  \frac{d}{dz}\frac1{N}\log\det\left( \I_{n_k} + c_ke_k(-z)\T_k \right) &= - z \cdot e_k'(-z)\cdot \delta_k(-z) \\
  \frac{d}{dz}\left( z\sum_{k=1}^K \delta_k(-z)e_k(-z)\right) &= \sum_{k=1}^K \delta_k(-z)e_k(-z) - z \sum_{k=1}^K \delta_k'(-z)\cdot e_k(-z)+\delta_k(-z)\cdot e_k'(-z).
\end{align*}

Combining the last three lines, we have
\begin{align*}
  &\sum_{k=1}^K \delta_k(-z)e_k(-z) = \nonumber \\ &\frac{d}{dz}\left[-\frac1N\log\det\left(\I_N + \sum_{k=1}^K \delta_k(-z) \R_k \right) - \sum_{k=1}^K \frac1{N}\log\det\left( \I_{n_k} + c_ke_k(-z)\T_k \right) + z\sum_{k=1}^K \delta_k(-z)e_k(-z) \right],
\end{align*}
which after integration leads to
\begin{align}
  \label{eq:capadelta}
  &\int_z^{+\infty} \left(\frac1w-m_N^\circ(-w) \right)dw = \nonumber \\ &\frac1N\log\det\left(\I_N + \sum_{k=1}^K \delta_k(-z) \R_k \right) + \sum_{k=1}^K \frac1{N}\log\det\left( \I_{n_k} + c_ke_k(-z)\T_k \right) - z\sum_{k=1}^K \delta_k(-z)e_k(-z),
\end{align}
which is exactly the right-hand side of \eqref{eq:th2V2}.

\subsection{Proof of \eqref{eq:proof22}}
Consider now the existence of a nonrandom $\alpha$ and for each $N$ a 
non-negative integer $r_N$ for which 
\begin{equation*}
\max_{i\leq K}\max(\lambda_{r_N+1}^{\T_i},\lambda_{r_N+1}^{\R_i})\leq \alpha
\end{equation*}
(eigenvalues also arranged in non-increasing order).
Then for each $i$
\begin{align*}
  \lambda_{2r_N+1}^{\R_i^\oh\X_i\T_i\X_i^\herm\R_i^\oh} &= (s_{2r_N+1}^{\R_i^\oh\X_i\T_i^\oh})^2 \\ 
& \leq \alpha^2\|\X_i\X_i^\herm\|
\end{align*}
and then we have, from Lemma \ref{le:th2},
\begin{equation*}
\lambda_{2Kr_N+1}^{\B_N}\leq \alpha^2(\|\X_1\X_1^\herm\|+\cdots+\|\X_K\X_K^\herm\|).
\end{equation*}

We can in fact consider that the spectral norms of the $\X_i$ are bounded in the limit. Either Gaussian assumptions on the components, or finite fourth moment, but all coming from doubly infinite arrays (remember though that we need the right-unitary invariance structure of $\X_i$).
Because of assumption \ref{item:ba0} in Corollary \ref{co:1}, we can, by enlarging the sample space, assume each $\X_i$ is embedded in an $N\times n'_i$ matrix $\X_i'$, where
$N/n_i'\to a$ as $N\to\infty$. Then, with probability one (see e.g. \cite{SIL06}), 
\begin{align}
  \label{eq:25}
\limsup_N\lambda_{2Kr_N+1}^{\B_N} &\leq \limsup_N\alpha^2 (\|\X_1'{\X_1'}^\herm\|+\cdots+\|\X_K'{\X_K'}^\herm\|)  \nonumber \\
&\leq \alpha^2\frac{Kb}{a}(1+\sqrt{a})^2.
\end{align}
Let $a^\circ $ be any real greater than $\alpha^2K\frac{b}{a}(1+\sqrt{a})^2$.  

Since $\S=0$ here, it follows as in \cite{SIL95} that $\{F^{\B_n}\}$ is almost surely tight. Let $F^\circ _N$ denote the distribution function having Stieltjes transform $m_N^\circ$, and let $f$ on $[0,\infty)$ be a continuous function. Then the function
\begin{equation*}
f_{a^\circ }(x)=
\left\{
\begin{array}{ll}
  f(x) &,\ x\leq a^\circ \\ f(a^\circ ) &,\ x>a^\circ 
\end{array}
\right.
\end{equation*}
is bounded and continuous. Therefore, with probability $1$,
\begin{equation*}
  \int f_{a^\circ }(x)dF_N(x)-\int f_{a^\circ }(x)dF^\circ _N(x)\to0,
\end{equation*}
as $N\to\infty$.

Suppose now $r_N=o(N)$. Then, since almost surely there are at most $2Kr_N$ eigenvalues greater than $a^\circ $ for all $N$ large, any converging subsequence of $\{F_N^\circ \}$ must have some mass lying on $[0,a^\circ ]$. This implies, with probability $1$,
\begin{equation*}
\frac1N\sum_{\lambda_i\leq a^\circ }f(\lambda_i)-\int_{[0,a^\circ ]}f(x)dF^\circ _N(x)\to0,
\end{equation*}
as $N\to\infty$.

Let $b_N$ be a bound on the spectral norms of the $\T_i$ and $\R_i$. Then
\begin{equation}
  \label{eq:26}
\|\B_n\|\leq b_N^2(\|\X_1'{\X_1'}^\herm\|+\cdots+\|\X_K'{\X_K'}^\herm\|).
\end{equation}

Fix a number $\beta>\frac{Kb}{a}(1+\sqrt{a})^2$, and let $a_N=b_N^2\beta$.
Suppose also that $f$ is increasing and that $f(a_N)r_N=o(N)$. Then
\begin{equation*}
\int f(x)dF^{\B_n}(x)-\frac1N\sum_{\lambda_i\leq a^\circ }f(\lambda_i)\to0,
\end{equation*}
almost surely, as $N\to\infty$. Therefore, with probability $1$,
\begin{equation*}
\int f(x)dF_N(x)-\int_{[0,a^\circ ]}f(x)dF^\circ _N(x)\to0,
\end{equation*}
as $N\to\infty$.

For any $N$ we consider, for $j=1,2,\ldots$, the $jN\times jN$ matrix $\B_{N,j}$ formed, as before, from block diagonal matrices and $jN\times jn_i$ matrices of i.i.d. variables. Then
with probability $1$, $F^{\B_{N,j}}$ converges weakly to $F_N^\circ $ as $j\to\infty$.
Properties on the eigenvalues of $\B_{N,j}$ will thus yield properties of
$F_N^\circ $.

By considering the bound on $\|\B_{n,j}\|$ analogous to \eqref{eq:26}, we must have $F_N^\circ (a_N)=1$ for all $N$ large. 

Similar to \eqref{eq:25} we see that, with probability $1$
\begin{equation*}
\limsup_j\lambda_{2Kjr_N+1}^{\B_{N,j}}\leq a^2((1+\sqrt{c_1})^2+\cdots+ (1+\sqrt{c_K})^2),
\end{equation*}
this latter number being less than $a^\circ $ for all $N$ large.   

At this point we will use the fact that for probability measures $P_N$, $P$ on $\RR$ with $P_N$ converging weakly to $P$, we have (see e.g. \cite{BIL68}) 
\begin{equation*}
\liminf_NP_N(G)\ge P(G)
\end{equation*}
for any open set $G$. Thus, with $G=(a^\circ ,\infty)$ we see that, with probability $1$, for all $N$ large 
\begin{align*}
  F^\circ _N((a^\circ ,\infty)) &= 1-F^\circ _N(a^\circ )\leq\liminf_jF^{\B_{N,j}}((a^\circ ,\infty)) \\ 
  &\leq 2Kr_N/N.
\end{align*}

Therefore, for all $N$ large
\begin{equation*}
\int_{(a^\circ ,\infty)}f(x)dF_N^\circ (x)\leq f(a_N)2Kr_N/N\to0,
\end{equation*}
as $N\to\infty$.

Therefore, we conclude that, $\int f(x)dF^\circ _N(x)$ is bounded, 
and with probability $1$
\begin{equation*}
\int f(x)dF_N(x)-\int f(x)dF^\circ _N(x)\to0,
\end{equation*}
as $N\to\infty$. This concludes the proof.

\section{Proof of Proposition \ref{prop:concavity}}
\label{app:concavity}
The proof stems from the following result,
\begin{proposition}
  \label{prop:concavf}
  $f(\P_1,\ldots,\P_K)$ is a strictly concave matrix in the Hermitian nonnegative definite matrices $\P_1,\ldots,\P_K$, if and only if, for any couples $(\P_{1_a},\P_{1_b}), \ldots,(\P_{K_a},\P_{K_b})$ of Hermitian nonnegative definite matrices, the function
\begin{equation*}
  \phi(\lambda) = f\left(\lambda\P_{1_a}+(1-\lambda)\P_{1_b},\ldots,\lambda\P_{K_a}+(1-\lambda)\P_{K_b} \right)
\end{equation*}
is strictly concave.
\end{proposition}
Denote
\begin{equation*}
  \bar{\mathcal V}^\circ_N(\lambda) = \mathcal V^\circ_N(\lambda\P_{1_a}+(1-\lambda)\P_{1_b},\ldots,\lambda\P_{|\mathcal S|_a}+(1-\lambda)\P_{|\mathcal S|_b})
\end{equation*}
and consider a set $(\delta_k,e_k,\P_{i_1},\ldots,\P_{i_{|\mathcal S|}})$ which satisfies the system of equations \eqref{eq:Idelta}-\eqref{eq:deltai}. Then, from remark \eqref{eq:dVdelta} and \eqref{eq:dVe},
\begin{align*}
  \frac{d \bar{\mathcal V}^\circ_N}{d \lambda} &= \sum_{k\in \mathcal S}\frac{\partial \bar{V}}{\partial \delta_k} \frac{\partial \delta_k}{\partial \lambda} + \frac{\partial \bar{V}}{\partial e_k} \frac{\partial e_k}{\partial \lambda} + \frac{\partial \bar{V}}{\partial \lambda} \\
  &= \frac{\partial \bar{V}}{\partial \lambda},
\end{align*}
where 
\begin{equation}
  \bar{V}: (\delta_1,\ldots,\delta_{|\mathcal S|},e_1,\ldots,e_{|\mathcal S|},\lambda) \mapsto \bar{\mathcal V}^\circ_N(\lambda).
\end{equation}

Mere derivations of $\bar{V}$ lead then to
\begin{equation*}
  \frac{\partial^2\bar{V}}{\partial \lambda^2} = - \sum_{i\in \mathcal S} (c_i^2e_i^2)\frac1N\tr\left(\I + c_ie_i\R_i\P_i\right)^{-2}(\R_i(\P_{i_a}-\P_{i_b}))^2.
\end{equation*}
Since $e_i> 0$ on the strictly negative real axis, if any of the $\R_i$'s is positive definite, then, for all nonnegative definite couples $(\P_{i_a},\P_{i_b})$, such that $\P_{i_a}\neq \P_{i_b}$, $\bar{\mathcal V}_N''<0$. Then, from Proposition \ref{prop:concavf}, the deterministic approximate $\mathcal V_N^\circ$ is strictly concave in $\P_1,\ldots,\P_{|\mathcal S|}$ if any of the $\R_i$ matrices is invertible.

\section{Proof of Proposition \ref{prop:wf}}
\label{ap:wf}
  The proof of Proposition \ref{prop:wf} recalls the proof from \cite{DUM10}, Proposition 5. 
  Let us define the functions  
  \begin{align}
    \label{eq:Idelta}
    \mathcal V_N^\circ(\P_1,\ldots,\P_{|\mathcal S|}) &= \sum_{k\in \mathcal S} \frac1{N}\log\det\left( \I_{n_k} + c_ke_k\R_k\P_k \right) \nonumber \\ &+ \frac1N\log\det\left(\I_N + \sum_{k\in \mathcal S} \delta_k \T_k \right) \nonumber \\
  &- \sigma^2\sum_{k=1}^K \delta_k(-\sigma^2)e_k(-\sigma^2),
  \end{align}
where
\begin{align}
  \label{eq:ei}
  e_i = e_i(\P_1,\ldots,\P_{|\mathcal S|}) &= \frac1N \tr \T_i\left(\sigma^2 \left[\I_N + \sum_{k\in \mathcal S}\delta_k\T_k\right] \right)^{-1} \\
  \label{eq:deltai}
  \delta_i = \delta_i(\P_1,\ldots,\P_{|\mathcal S|}) &= \frac1{n_i} \tr \R_i\P_i \left(\sigma^2 \left[\I_{n_i} + c_ie_i(z)\R_i\P_i \right] \right)^{-1}
\end{align}
and $V:(\P_1,\ldots,\P_{|\mathcal S|},\delta_1,\ldots,\delta_{|\mathcal S|},e_1,\ldots,e_{|\mathcal S|})\mapsto \mathcal V_N^\circ(\P_1,\ldots,\P_{|\mathcal S|})$. Then we need only prove that, for all $k\in \mathcal S$, 
\begin{align*}
  \frac{\partial V}{\partial \delta_k}(\P_1,\ldots,\P_{|\mathcal S|},\delta^\circ_1,\ldots,\delta^\circ_{|\mathcal S|},e^\circ_1,\ldots,e^\circ_{|\mathcal S|}) &= 0 \\
  \frac{\partial V}{\partial e_k}(\P_1,\ldots,\P_{|\mathcal S|},\delta^\circ_1,\ldots,\delta^\circ_{|\mathcal S|},e^\circ_1,\ldots,e^\circ_{|\mathcal S|}) &= 0 .
\end{align*}
Remark then that
\begin{align}
  \label{eq:dVdelta}
  \frac{\partial V}{\partial \delta_k}(\P_1,\ldots,\P_{|\mathcal S|},\delta_1,\ldots,\delta_{|\mathcal S|},e_1,\ldots,e_{|\mathcal S|}) &= \frac1N\tr\left[\left(\I+\sum_{i\in \mathcal S}\delta_i \T_i\right)^{-1}\T_k \right] - \sigma^2e_k \\
  \label{eq:dVe}
\frac{\partial V}{\partial e_k}(\P_1,\ldots,\P_{|\mathcal S|},\delta_1,\ldots,\delta_{|\mathcal S|},e_1,\ldots,e_{|\mathcal S|}) &= c_k\frac1N\tr\left[\left(\I + c_ke_k \R_i\P_i\right)^{-1}\R_k\P_k \right] - \sigma^2\delta_k,
\end{align}
both being null whenever, for all $k$, $e_k=e_k(-\sigma^2,\P_1,\ldots,\P_{|\mathcal S|})$ and $\delta_k=\delta_k(-\sigma^2,\P_1,\ldots,\P_{|\mathcal S|})$, which is true in particular for the unique power optimal solution $\P_1^\circ,\ldots,\P_{|\mathcal S|}^\circ$ whenever $e_k=e_k^\circ$ and $\delta_k=\delta_k^\circ$.

When, for all $k$, $e_k=e_k^\circ$, $\delta_k=\delta_k^\circ$, the maximum of $V$ over the $\P_k$ is then obtained by maximising the expressions $\log\det(\I_{n_k}+c_ke_k^\circ \R_k\P_k)$ over $\P_k$. 
From the inequality (see e.g. \cite{TEL99})
\begin{equation*}
	\det(\I_{n_k}+c_ke_k^\circ\R_k\P_k) \leq \prod_{i=1}^{n_k} \left(\I_{n_k} + c_ke_k^\circ\R_k\P_k\right)_{ii},
\end{equation*}
where, only here, we denote $(\X)_{ii}$ the entry $(i,i)$ of matrix $\X$. The equality is obtained if and only if $\I_{n_k}+c_ke_k^\circ\R_k\P_k$ is diagonal. The equality case arises for $\P_k$ and $\R_k=\U_k\D_k\U_k^\herm$ co-diagonalizable. In this case, denoting $\P_k=\U_k\Q_k\U_k^\herm$, the entries of $\Q_k$, constrained by $\frac1{n_k}\tr(\Q_k)=P_k$ are solutions of the classical optimisation problem under constraint,
\begin{align*}
  \sup_{\substack{\Q_k \\ \frac1{n_k}\tr(\Q_k)\leq P_k} }\log\det\left(\I_{n_k}+ c_ke_k^\circ \Q_k\D_k\right),
\end{align*}
whose solution is given by the classical water-filling algorithm. Hence \eqref{eq:qki}. 

\section{Proof of Proposition \ref{prop:convergence}}
\label{ap:fp}
The convergence of the fixed-point algorithm follows the same line of proof as the uniqueness in Section \ref{sec:proof_K}. We prove the convergence for $z\in\CC^+$, although this can be easily generalised. If one considers the difference $\e^{n+1}-\e^n$, where $\e^n=(e^n_1,\ldots,e^n_K)$, instead of $\e^\circ -\underline \e^\circ $, the same development as in Section \ref{sec:proof_K} leads to
\begin{equation*}
	\e^{n+1}-\e^n = \A_n(\e^n-\e^{n-1})
\end{equation*}
for $n\geq 1$, where $\A_n$ is defined, similarly as in \eqref{eq:aij}, as $\A_n=(a^n_{ij})$, with $a_{ij}^n$ defined by
\begin{equation*}
	a^n_{ij}=\frac1N\tr \R_i{\D_{n-1}}^{-1}\R_j{\D_n}^{-1}c_j\int\frac{\tau^2}{(1+c_j\tau e_j^{n-1})(1+c_j\tau e_j^n)}dF^{\T_j}(\tau),
\end{equation*}
where $\D_n$ is $\D$ for $e_j(z)$ replaced by $e_j^n(z)$.

From Cauchy-Schwarz inequality, and the different bounds on the $\D_n$, $\R_k$ and $\T_k$ matrices used so far, we have
\begin{equation*}
	a^n_{ij} \leq \frac{|z|^2c_j}{v^4} \frac{\log N^4}N,
\end{equation*}
with $v=\Im[z]$. Denoting $c_0=\max(c_j)$, we then have that
\begin{equation*}
	\max_j\left(e^{n+1}_j - e^n_j \right) < K\frac{|z|^2c_0}{v^4}\leq \frac{\log N^4}N \max_j\left(e^{n}_j - e^{n-1}_j \right).
\end{equation*}

Let $0<\varepsilon<1$, and take now a countable set $\{z_1,z_2,\ldots\}$, $v_k=\Im[z_k]$, such that $K\frac{|z_k|^2c_0}{v_k^4}\frac{\log N^4}N<1-\varepsilon$ for all $z_k$ (this is possible by letting $v_k>0$ be large enough). On this countable set, the sequences $\{\e^n\}$ are therefore Cauchy sequences on $\CC^K$: they all converge. Since the $e^n_j$ are holomorphic and bounded on every compact set included in $\CC\setminus \RR^+$, from Vitali's convergence theorem \cite{TIT39}, the function $e^n_j(z)$ converges on such compact sets. Now, from the fact that we forced the initialisation step to be $e^0_j=-1/z$, $e^0_j$ is the Stieltjes transform of a distribution function at point $z$. It now suffices to verify that, if $e^n_j$ is the Stieltjes transform of a distribution function at point $z$, then so is $e^{n+1}_j$. This requires to verify that $z\in\CC^+,~e^n_j\in\CC^+$ implies $e^{n+1}_j\in\CC^+$, $z\in\CC^+,~ze^n_j\in\CC^+$ implies $ze_j^{n+1}\in\CC^+$, and $\lim_{y\to \infty} -ye^n_j(iy)<\infty$ implies that $\lim_{y\to \infty} -ye^n_j(iy)<\infty$. This follows directly from the definition of $e^n_j$. From the dominated convergence theorem, we then also have that the limit of $e^n_j$ is a Stieltjes transform that is solution to \eqref{eq:th1en}. From the uniqueness of the Stieltjes transform, solution to \eqref{eq:th1en} (this follows from the pointwise uniqueness on $\CC^+$ and the fact that the Stieltjes transform is holomorphic on all compact sets of $\CC\setminus\RR^+$), we then have that $e^n_j$ converges for all $j$ and $z\in\CC\setminus\RR^+$, if $e^0_j$ is initialised at a Stieltjes transform. 

\section{Useful Lemmas}
\label{sec:app1}
In this section, we gather most of the known or new lemmas which are needed in various places in the proof of Appendices \ref{ap:th1}-\ref{ap:fp}.

The statements in the following Lemma are well-known
\begin{lemma}
  \label{le:1}
  \begin{enumerate}
    \item \label{le:1a} For rectangular matrices $\A$, $\B$ of the same size,
      \begin{equation*}
	\rank(\A + \B) \leq \rank(\A)+\rank(\B);
      \end{equation*}
    \item \label{le:1b} For rectangular matrices $\A$, $\B$ for which $\A\B$ is defined,
      \begin{equation*}
	\rank(\A\B)\leq \min (\rank(\A),\rank(\B));
      \end{equation*}
    \item \label{le:1c} For rectangular $\A$, $\rank(\A)$ is less than the number of non-zero entries of $\A$.
  \end{enumerate}
\end{lemma}

\begin{lemma}
  (Lemma 2.4 of \cite{SIL95}) For $N\times N$ Hermitian matrices $\A$ and $\B$,
  \begin{equation*}
    \Vert F^{\A}-F^{\B}\Vert \leq \frac{1}{N}\rank(\A-\B).
  \end{equation*}
\end{lemma}

From these two lemmas we get the following.
\begin{lemma}
  \label{le:3}
Let $\S$, $\A$, $\overline \A$, be Hermitian $N\times N$, $\Q$, $\overline \Q$ both $N\times n$, and $\B$, $\overline \B$ both Hermitian $n\times n$. Then
\begin{enumerate}
  \item
  \label{le:3a}
\begin{equation*}
\Vert F^{\S+\A\Q\B\Q^\herm \A}-F^{\S+\A\overline \Q\B\overline \Q^\herm \A}\Vert \leq\frac2N\rank(\Q-\overline \Q);
\end{equation*}
\item 
  \label{le:3b}
  \begin{equation*}
\Vert F^{\S+\A\Q\B\Q^\herm \A}-F^{\S+\overline \A\Q\B\Q^\herm\overline \A}\Vert\leq\frac2N\rank(\A-\overline \A);
\end{equation*}
\item
  \label{le:3c}
  \begin{equation*}
\Vert F^{\S+\A\Q\B\Q^\herm \A}-F^{\S+\A\Q\overline \B\Q^\herm \A}\Vert\leq\frac1N\rank(\B-\overline \B).
\end{equation*}
\end{enumerate}
\end{lemma}

\begin{lemma}
  \label{le:4}
For $N\times N$ $\A$, $\tau\in\CC$ and $\rv\in\CC^N$ for which $\A$ and $\A+\tau \rv\rv^\herm$ are invertible,
\begin{equation*}
\rv^\herm(\A+\tau \rv\rv^\herm)^{-1}=\frac1{1+\tau \rv^\herm \A^{-1}\rv}\rv^\herm \A^{-1}.
\end{equation*}
\end{lemma}
This result follows from $\rv^\herm \A^{-1}(\A+\tau \rv\rv^\herm)=(1+\tau \rv^\herm \A^{-1}\rv)\rv^\herm$.

Moreover, we recall Lemma 2.6 of \cite{SIL95}
\begin{lemma}
  \label{le:2.6}
Let $z\in\CC^+$ with $v=\Im[z]$, $\A$ and $\B$ $N\times N$ with $\B$ Hermitian, and $\rv\in \CC^N$. Then
\begin{equation*}
\left|\tr\left((\B-z\I_N)^{-1}-(\B+\rv\rv^\herm-z\I_N)^{-1}\right)\A\right|=
\left|\frac{\rv^\herm(\B-z\I_N)^{-1}\A(\B-z\I_N)^{-1}\rv}
{1+\rv^\herm(\B-z\I_N)^{-1}\rv}\right|\leq\frac{\|\A\|}v.
\end{equation*}
If $z<0$, we also have
\begin{equation*}
  \left|\tr\left((\B-z\I_N)^{-1}-(\B+\rv\rv^\herm-z\I_N)^{-1}\right)\A\right|\leq\frac{\|\A\|}{|z|}.
\end{equation*}
\end{lemma}

From Lemma 2.2 of \cite{SHO70}, and Theorems A.2, A.4, A.5 of \cite{KRE97}, we have the following
\begin{lemma}
  \label{le:5}
If $f$ is analytic on $\CC^+$, both $f(z)$ and $zf(z)$ map $\CC^+$ into $\CC^+$, and there exists a $\theta\in(0,\pi/2)$ for which $zf(z)\to c$, finite, as $z\to\infty$ restricted to $\{w\in \CC:\theta<\arg w < \pi-\theta\}$, then $c<0$ and $f$ is the Stieltjes transform of a measure on the nonnegative reals with total mass $-c$.
\end{lemma}

Also, from \cite{SIL95}, we need
\begin{lemma}
  \label{le:13}
  Let $\y=(y_1,\ldots,y_N)^\trans$ with the $y_i$'s i.i.d. such that $\EE y_1=0$, $\EE |y_1|^2=1$ and $y_1\leq \log N$, and $\A$ an $N\times N$ matrix independent of $\y$, then
  \begin{equation*}
    \EE |\y^\herm \A \y - \tr \A|^6\leq K\Vert \A \Vert^6N^3\log^{12} N,
  \end{equation*}
  where $K$ does not depend on $N$, $\A$, nor on the distribution of $y_1$.
\end{lemma}

Additionally, we need 
\begin{lemma}
  \label{le:6}
  Let $\D=\A+i\B+iv\I$, where $\A$, $\B$ are $N\times N$ Hermitian,
$B$ is also positive semi-definite, and $v>0$. Then $\|\D^{-1}\|\leq v^{-1}$.
\end{lemma}
\begin{IEEEproof}
We have $\D\D^\herm=(\A+i\B)(\A-i\B)+v^2\I+2v\B$. Therefore the eigenvalues of
$\D\D^\herm$ are greater or equal to $v^2$, which implies the singular values of $\D$ are greater or equal to $v$, so that the singular values of $\D^{-1}$ are less or equal to $v^{-1}$.  We therefore get our result.
\end{IEEEproof}

From Theorem 2.1 of \cite{SEN81},
\begin{lemma}
  \label{le:7} 
  Let $\rho(\CCC)$ denote the spectral radius of the $N\times N$ matrix $\CCC$ (the largest of the absolute values of the eigenvalues of $\C$). If $\x,\b\in\RR^N$ with the components of $\CCC$, $\x$, and $\b$ all positive, then the equation $\x=\CCC\x+\b$ implies $\rho(\CCC)<1$.
\end{lemma}

From Theorem 8.1.18 of \cite{HOR85},
\begin{lemma}
  \label{le:8}
Suppose $\A=(a_{ij})$ and $\B=(b_{ij})$
are $N\times N$ with $b_{ij}$ nonnegative and $|a_{ij}|\leq b_{ij}$. Then 
\begin{equation*}
  \rho(\A)\leq\rho((|a_{ij}|))\leq\rho(\B).
\end{equation*}
\end{lemma}

Also, from Lemma 5.7.9 of \cite{HOR91},
\begin{lemma} 
  \label{le:9}
  Let  $\A=(a_{ij})$ and $\B=(b_{ij})$ be $N\times N$ with $a_{ij}$, $b_{ij}$ nonnegative. Then
  \begin{equation*}
\rho((a_{ij}^\oh b_{ij}^\oh))\leq(\rho(\A))^\oh(\rho(\B))^\oh.
\end{equation*}
\end{lemma}

And, Theorems 8.2.2 and 8.3.1 of \cite{HOR85},
\begin{lemma}
  \label{le:10}
If $\CCC$ is a square matrix with nonnegative entries, then $\rho(\CCC)$ is an eigenvalue of $\CCC$ having an eigenvector $\x$ with nonnegative entries. Moreover, if the entries of $\CCC$ are all positive, then $\rho(\CCC)>0$  and the entries of $\x$ are all positive.
\end{lemma}

From \cite{HOR91}, we also need Theorem 6.1.1,
\begin{lemma} {\it Gersgorin's Theorem}
  \label{le:11}
  All the eigenvalues of an $N\times N$  matrix $\A=(a_{ij})$ lie in the 
union of the $N$ disks in the complex plane, the $i^{th}$ disk
having center $a_{ii}$ and radius $\sum_{j\neq i}|a_{ij}|$.
\end{lemma}

Theorem 3.42 of \cite{TIT39},
\begin{lemma}
  \label{le:12}
  {\it Rouche's Theorem}
  If $f(z)$ and $g(z)$ are analytic inside and on a closed contour $C$ of the complex plane,
and $|g(z)|<|f(z)|$ on $C$, then $f(z)$ and $f(z)+g(z)$ have the same number of zeros inside $C$.
\end{lemma}

In order to prove Theorem \ref{th:2}, we also need, from \cite{FAN51}
\begin{lemma}
  \label{le:th2}
Consider a rectangular matrix $\A$ and let $s_i^\A$ denote the $i^{th}$ largest
singular value of $\A$, with $s_i^\A=0$ whenever $i>\text{rank}(A)$. Let 
$m$, $n$ be arbitrary non-negative integers. Then for $\A$, $\B$ rectangular
of the same size
\begin{equation*}
s_{m+n+1}^{\A+\B}\leq s_{m+1}^\A+s_{n+1}^\B,
\end{equation*}
and for $\A$, $\B$ rectangular for which $\A\B$ is defined
\begin{equation*}
s_{m+n+1}^{AB}\leq s_{m+1}^As_{n+1}^B.
\end{equation*}
As a corollary, for any integer $r\ge0$ and rectangular matrices $\A_1,\ldots,\A_K$, all of the same size,
\begin{equation*}
s_{Kr+1}^{\A_1+\cdots+\A_K}\leq s_{r+1}^{\A_1}+\cdots+s_{r+1}^{\A_K}.
\end{equation*}
\end{lemma}

\bibliography{tutorial_RMT/IEEEconf,tutorial_RMT/IEEEabrv,tutorial_RMT/tutorial_RMT}

\begin{IEEEbiography}[{\includegraphics[width=1in,height=1.25in]{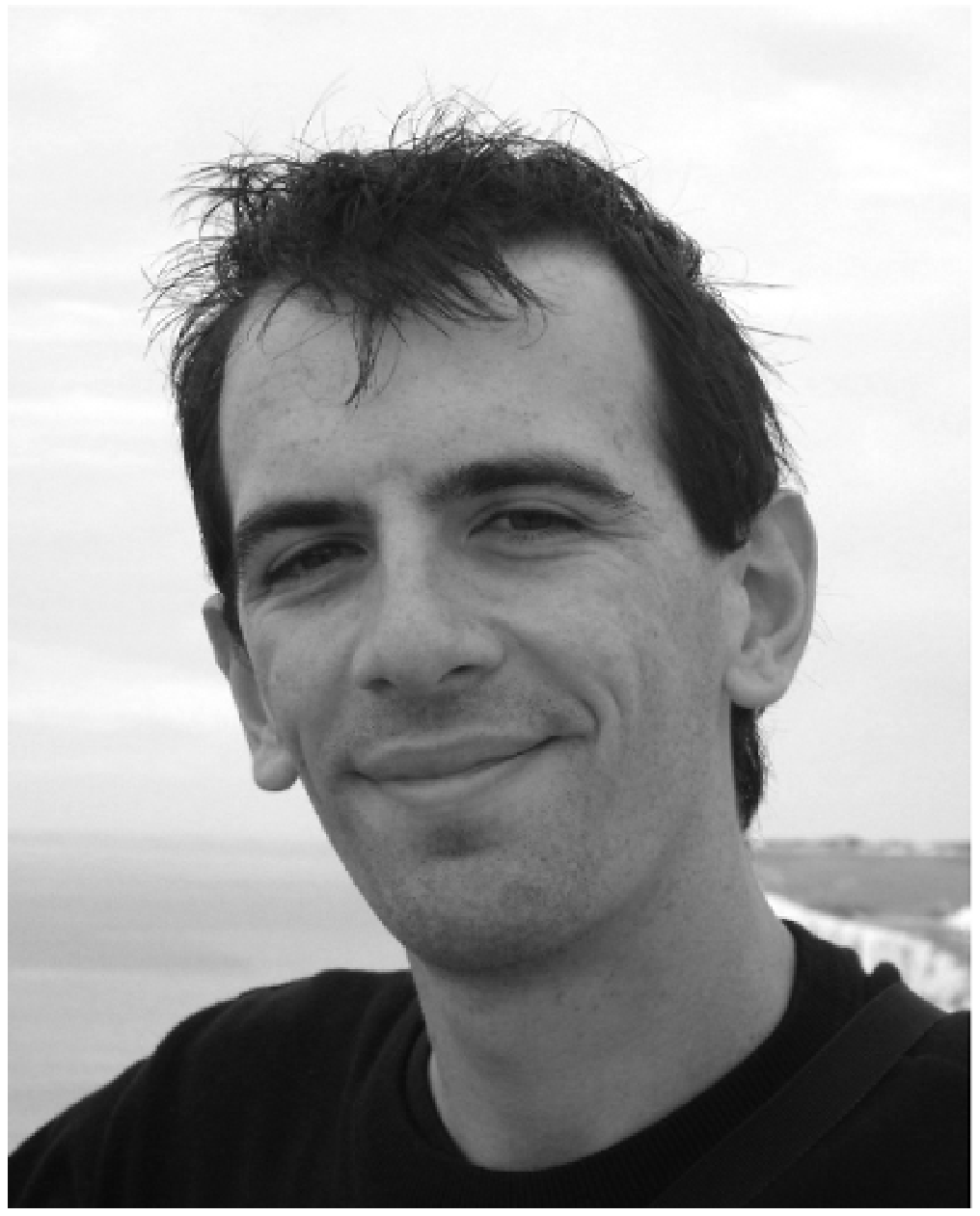}}]{Romain Couillet}
was born in Abbeville, France. He received his Msc. in Mobile Communications at the Eurecom Institute, France in 2007. He received his Msc. in Communication Systems in Telecom ParisTech, France in 2007. In September 2007, he joined ST-Ericsson (formerly NXP Semiconductors, founded by Philips). At ST-Ericsson, he works as an Algorithm Development Engineer on the Long Term Evolution Advanced (LTE-A) project. In parallel to his position at ST-Ericsson, he is currently a PhD student at Supélec, France. His research topics include mobile communications, multi-users multi-antenna detection, cognitive radio cognitive, Bayesian probability and random matrix theory. He is the recipient of the ValueTools best student paper award, 2008.
\end{IEEEbiography}

\begin{IEEEbiography}[{\includegraphics[width=1in,height=1.25in]{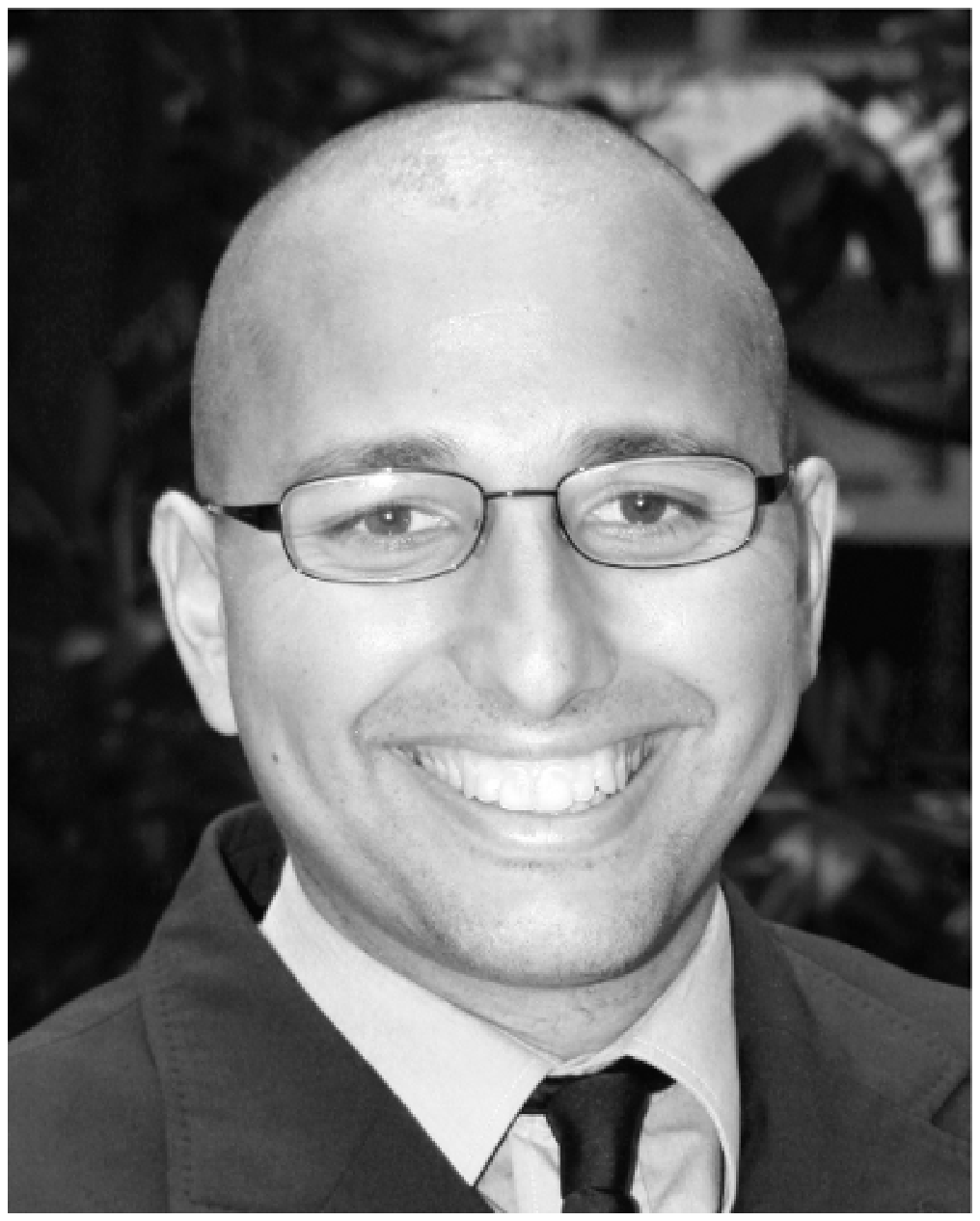}}]{M\'erouane Debbah}
was born in Madrid, Spain. He entered the Ecole Normale Supérieure de Cachan (France) in 1996 where he received his M.Sc and Ph.D. degrees respectively in 1999 and 2002. From 1999 to 2002, he worked for Motorola Labs on Wireless Local Area Networks and prospective fourth generation systems. From 2002 until 2003, he was appointed Senior Researcher at the Vienna Research Center for Telecommunications (FTW) (Vienna, Austria). From 2003 until 2007, he joined the Mobile Communications de-partment of the Institut Eurecom (Sophia Antipolis, France) as an Assistant Professor. He is presently a Professor at Supelec (Gif-sur-Yvette, France), holder of the Alcatel-Lucent Chair on Flexible Radio. His research interests are in information theory, signal processing and wireless communications. Mérouane Debbah is the recipient of the ``Mario Boella'' prize award in 2005, the 2007 General Symposium IEEE GLOBECOM best paper award, the Wi-Opt 2009 best paper award, the2010  Newcom++ best paper award  as well as the Valuetools 2007,Valuetools 2008 and CrownCom2009 best student paper awards. He is a WWRF fellow. 
\end{IEEEbiography}

\begin{IEEEbiography}[{\includegraphics[width=1in,height=1.25in]{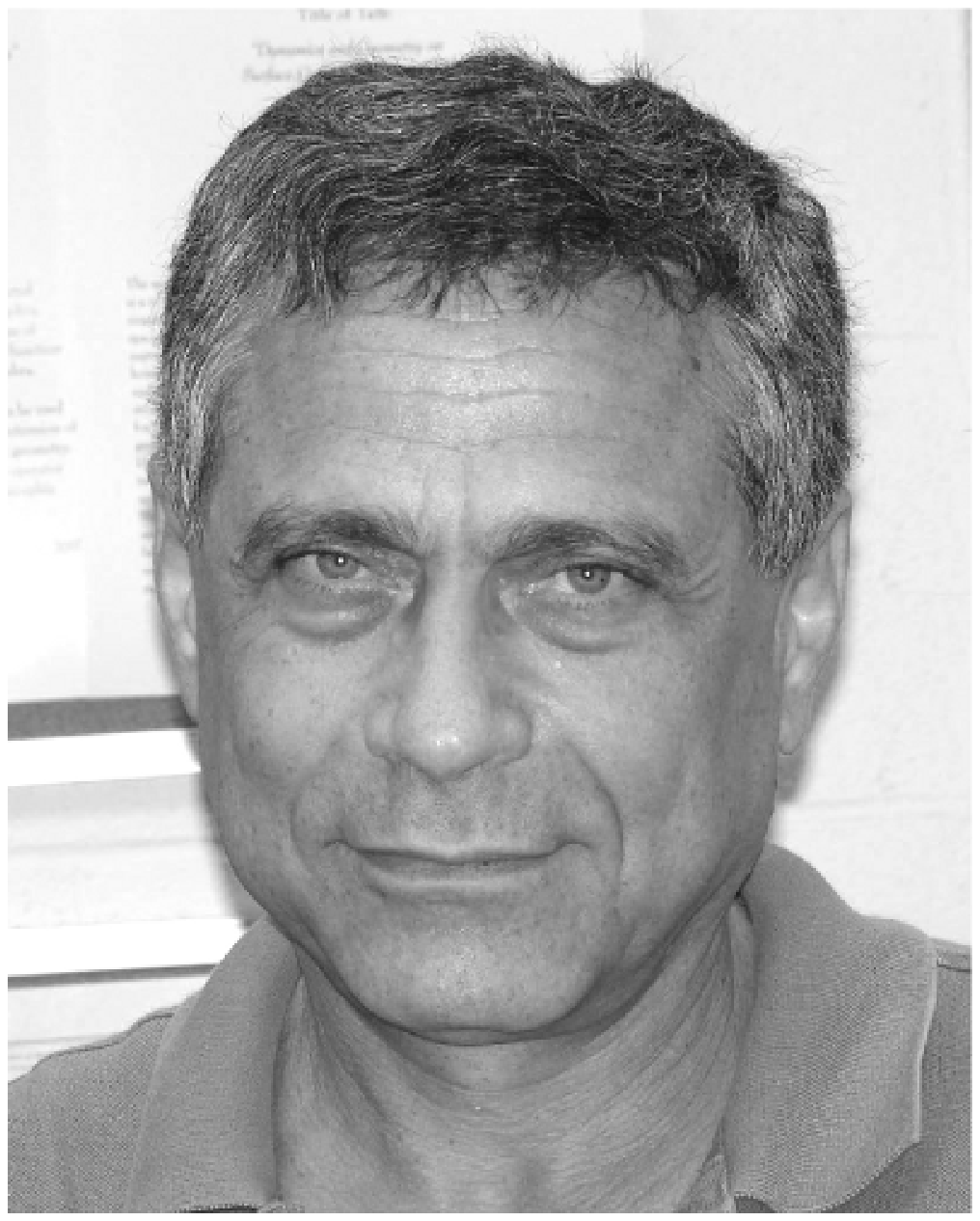}}]{Jack W. silverstein}
received the B.A. degree in mathematics from Hofstra University, Hempstead, NY, in 1971 and the M.S. and Ph.D. degrees in applied mathematics from Brown University, Providence, RI, in 1973 and 1975, respectively.  After postdoctoring and teaching at Brown, he began in 1978 a tenure track position in the Department of Mathematics at North Carolina State University, Raleigh, where he has been Professor since 1994. His research interests are in probability theory with emphasis on the spectral behavior of large-dimensional random matrices. Prof. Silverstein was elected Fellow of the Institute of Mathematical Statistics in 2007. 
\end{IEEEbiography}

\end{document}